\documentclass[pre,floats,superscriptaddress]{revtex4}

\usepackage[dvips]{graphicx}
\usepackage{amssymb}
\usepackage{amsmath,bm}
\usepackage{psfrag}
\usepackage{epsfig}
\usepackage{float}

\begin{document}
\title{Step Growth and Meandering in a Precursor-Mediated Epitaxy with Anisotropic Attachment Kinetics and Terrace Diffusion}

\author{Mikhail Khenner}
\affiliation{Department of Mathematics and Applied Physics Institute, Western Kentucky University, Bowling Green, KY 42101}

\begin{center}
\small{{\it Journal information: Mathematical Modelling of Natural Phenomena Vol. 10, No. 4, 2015, pp.96-109; DOI: 10.1051/mmnp/201510406}}
\end{center}

\begin{abstract}
Step meandering instability in a Burton-Cabrera-Frank (BCF)-type model for the growth of an isolated, atomically high step on a crystal surface is analyzed.
It is assumed that the growth  is sustained by the molecular precursors deposition 
on a terrace and their decomposition into atomic constituents; both processes are explicitly modeled. 
A strongly nonlinear evolution PDE for the shape of the step is derived in the long-wave limit and without assuming smallness of the amplitude; this equation may be 
transformed into a convective Cahn-Hilliard-type PDE for the step slope.
Meandering is 
studied as a function of the precursors diffusivity and of the desorption rates of the precursors and adatoms. 
Several important features are identified, such as: the interrupted coarsening, ``facet" bunching, and the lateral drift of the step perturbations (a traveling wave) when
the terrace diffusion is anisotropic.
The nonlinear drift introduces a disorder into the evolution of a step meander, which results in a pronounced oscillation of the step velocity, meander amplitude and lateral length scale
in the steady-state that emerged after the coarsening was interrupted. The mean values of these characteristics are also strongly affected by the drift. 


\end{abstract}

\maketitle


\setcounter{equation}{0}
\section{Introduction}
\label{Intro}

This paper investigates the dynamics of a crystal step (the terrace edge) in the conditions that mimic those
in the chemical vapor or beam epitaxy of thin films and its variants (the chemical vapor deposition, metal-organic
vapor phase epitaxy, etc.) In the chemical epitaxy the precursor molecules are deposited
onto a crystal surface which consists of alternating terraces and steps, undergo diffusion on the terraces 
and then decompose into the atomic constituents. These adatoms in turn undergo diffusion and then
are incorporated into the solid by attachment at the steps. 

Despite the abundance of the physico-mathematical models  of thwe molecular beam epitaxy (based on a partial differential equations, atomistic,
or multi-scale) - whereby there is a single deposited species and no chemical reaction - there is only a handful of models for the epitaxy of multiple atomic species with or without chemical
reaction effects.  For example, in Refs. \cite{CTT,CTT1} and \cite{W} the authors study the reaction-diffusion models for the dynamics of a monolayer, where the
reaction terms represent the adsorption-desorption and chemical processes, and the diffusion is affected by the nearest-neighbor attractive interactions between the deposited atoms and
also by the adatoms-substrate interactions. They find the emergence of self-assembled one-dimensional structures that may precipitate different types of film growth mechanisms.   
In Refs. \cite{CJ1,CJ2} an elaborate thermomechanical model for step-flow growth is analyzed, where the diffusing species are coupled through a chemical reaction whereby bulk molecules 
are crystallized from adatoms attaching to the step edges; the model ensures a configurational force balance at the steps through the generalization of the classical Gibbs-Thomson relation.  
And in Refs. \cite{GMPV,CPRDKZV} a simpler one-dimensional BCF\cite{BCF}-type model
is developed and applied for the analysis of step bunching. In this model the diffusion of adatoms is one-way
coupled - through the source term in the diffusion equation - to the diffusion of precursor molecules. The source term originates in decomposition
of the precursor, and the corresponding physical boundary conditions are formulated for both species at the step.  The latter model 
has its predecessor in the work by Pimpinelli \textit{et al.} \cite{PCGNT,PV}.

In this paper the model of Refs. \cite{GMPV,CPRDKZV} is extended
to two dimensions, where the step edge is a plane curve. The goal is to study how the interplay of the precursor and adatom diffusion, desorption,
precursor decomposition and adatom attachment to the step influence the step stability and
its growth. (Note that desorption of adatoms is neglected in Refs. \cite{GMPV,CPRDKZV}.)
Step meandering (lateral modulations) is the prominent feature in the experiments \cite{GMPV}. The meandering can be only studied theoretically using 
a two-dimensional model.

The first stage in such study is to consider processes such as diffusion on a single (lower) terrace
bordering the step (the one-sided model). This greatly simplifies derivations while retaining most
of the important physics, and thus we adopt such approach. Besides this simplification, the only
other simplification within the framework of BCF-type modeling is the neglect of the adatom diffusion along the step (the line diffusion). 
On the other hand, we include the rarely considered (due to the complexity of a treatment) 
diffusion anisotropy on a terrace, both for the precursors and adatoms; the anisotropy of line energy; and the anisotropic attachment kinetics. Thus we consider the full range of anisotropic effects, whose
importance was very recently re-emphasized in connection to the island growth on terraces \cite{MSL}.
It was also argued by same authors in Ref. \cite{MLKMS} that these effects, as well as the coupled dynamics
of the precursors and adatoms are important in epitaxial graphene growth on metals. We largely follow these
papers in the introduction and notations of the anisotropies. (For a more general treatment of the terrace diffusion anisotropy see 
Ref. \cite{DPKM}.) 

The derivation relies on a long-wave expansion. There is close physical and mathematical
analogy (noted by many authors and also in Ref. \cite{MLKMS}) between growth of a crystal into a hypercooled melt and the step growth within the one-sided model. Thus in developing the long-wave evolution PDE for the
step profile we pay close attention to the paper by Golovin \textit{et al.} \cite{GDN}, who
applied this framework to investigate faceting of the growing crystal surface. Our evolution PDE will be a generalization of the one they obtained. (It must be noted that we are concerned only with a weakly
anisotropic step energy, in contrast to a strongly anisotropic surface energy in Ref. \cite{GDN}.)

\setcounter{equation}{0}
\section{Problem formulation}
\label{Formulation}

We consider a morphological evolution of an unstable monoatomic step on a crystal surface. A step grows by the flux of adatoms from the lower terrace. The adatoms are the product of 
the precursors decomposition, and the latter are deposited on a terrace either by condensation from a vapor phase or by a molecular beam. (If the initially straight step is at $z=0$, then the lower terrace is the domain $z>0$.)

The governing equations of the model are the steady-state diffusion equations for the concentrations of the precursors (A) and adatoms (C) on the lower terrace, the mass conservation conditions at the terrace edge $z=h(x,t)$
(the step),  the Gibbs-Thomson boundary conditions for the concentrations at the step, and the boundary conditions for the concentrations on the lower terrace far from the step. The problem for the precursors reads \cite{GMPV,CPRDKZV}:
\begin{equation}
\nabla \cdot\left(D_a\nabla A\right) - \tau_a^{-1}A - \chi A = -F,
\label{diff_eq_A}
\end{equation}
\begin{equation}
z=h(x,t):\quad \left(D_a\nabla A\right)\cdot {\mathbf n} = \beta_a A, 
\nonumber
\end{equation}
\begin{equation}
z\rightarrow \infty:\quad A = \frac{F}{\tau_a^{-1} + \chi}.
\nonumber
\end{equation}
Here $D_a$ is the diffusion tensor, $\tau_a^{-1}$ and $\chi$ are the desorption and decomposition rates on a terrace, $F$ is the  deposition flux, $\beta_a$ is the kinetic coefficient, 
and ${\mathbf n}$ is the unit normal to the step pointing into lower terrace.  

The problem for the adatoms reads:
\begin{equation}
\nabla \cdot\left(D_c\nabla C\right) - \tau_c^{-1}C = -\chi A,
\label{diff_eq_C}
\end{equation}
\vspace{-0.4cm}
\[
z=h(x,t):\quad v_n \equiv h_t \cos{\theta} = \Omega\left(D_c\nabla C\right)\cdot {\mathbf n},
\]
\begin{equation}
C = C_{eq}\left( 1+ \frac{\Omega}{k_B\bar T}\tilde \beta_s(\theta)\kappa\right) + \tilde \beta_k(\theta) v_n,   \label{GT} 
\end{equation}
\begin{equation}
z\rightarrow \infty:\quad C\;\; \mbox{is bounded},
\nonumber
\end{equation}
where $D_c$ is the diffusion tensor, $\tau_c^{-1}$ is the desorption rate on a terrace, $v_n$ is the step normal velocity, $\Omega$ is the area occupied by an atom, $C_{eq}$ is the equilibrium concentration, $k_B\bar T$ is Boltzmann's factor, $\tilde \beta_s(\theta)$ and $\kappa$ are the step stiffness and the curvature, and $\tilde \beta_k(\theta) = \beta_k(\theta)/(k_+\Omega)$ is the 
kinetic coefficient (here $\beta_k(\theta)$ is a dimensionless anisotropy function of the reciprocal attachment coefficient, $1/k_+$ \cite{MSL}). Also $\theta$ is the angle of ${\mathbf n}$ 
with the $z$-axis.
The line diffusion is assumed insignificant and thus
this contribution is not included in Eq. (\ref{GT}).  The adatom diffusion is coupled to the precursors diffusion
through the term $-\chi A$ at the rhs of Eq. (\ref{diff_eq_C}). This term provides the continuous source of adatoms resulting from decomposition of the precursors; the same term 
appears with the opposite sign in Eq. (\ref{diff_eq_A}) \cite{CPRDKZV}. 

For the time being, we write the diffusion tensors as $D_a=\bar D_a \tilde D_a(\psi), D_c=\bar D_c \tilde D_c(\psi)$, where
$\bar D_a, \bar D_c$ are the magnitudes, and $\tilde D_a(\psi), \tilde D_c(\psi)$ are the dimensionless anisotropies \cite{MSL}.
The length scale is chosen equal to the characteristic distance that a precursor diffuses prior to decomposition:
$\ell = \sqrt{\bar D_a/\chi}$ \cite{CPRDKZV}. Choosing
$\ell^2/\bar D_a =1/\chi$ as the time scale, writing $A=\hat A/\Omega, C= C_{eq}+\hat C/\Omega$, results in the following dimensionless problems.
For the precursors:
\begin{equation}
\nabla \cdot\left(\tilde D_a\nabla \hat A\right) - t_a^{-1}\hat A = -f,
\label{diff_eq_A_ndim}
\end{equation}
\begin{equation}
z=h(x,t):\quad \left(\tilde D_a\nabla \hat A\right)\cdot {\mathbf n} = \hat \beta_a \hat A, \label{A_cond_h_ndim}
\end{equation}
\begin{equation}
z\rightarrow \infty:\quad \hat A = t_a f.
\label{infty_A_ndim}
\end{equation}
For the adatoms:
\begin{equation}
\nabla \cdot\left(\tilde D_c\nabla \hat C\right) - t_c^{-1}\hat C = -\hat \chi \hat A + g,
\label{diff_eq_C_ndim}
\end{equation}
\vspace{-0.4cm}
\begin{equation}
z=h(x,t):\quad \hat C = d_0 \beta_s(\theta) \kappa + \beta_0  \beta_k(\theta) \left(\tilde D_c\nabla \hat C\right)\cdot {\mathbf n},   \label{GT_ndim} 
\end{equation}
\begin{equation}
z\rightarrow \infty:\quad \hat C\;\; \mbox{is bounded}.
\label{infty_C_ndim}
\end{equation}

After the adatom concentration has been determined from Eqs. (\ref{diff_eq_C_ndim})-(\ref{infty_C_ndim}), the step profile dynamics is found from
\begin{equation}
h_t \cos{\theta} = \bar D\left(\tilde D_c\nabla \hat C_{|z=h(x,t)}\right)\cdot {\mathbf n}, \label{h_eq_ndim}
\end{equation}
where ${\mathbf n} = \left(-h_x \cos{\theta}, \cos{\theta}\right)$, and  $\cos{\theta} = \left(1+h_x^2\right)^{-1/2}$.
We used the same notations for the dimensionless $x,z,t,h,\kappa,\nabla,v_n$. The parameters are: $t_a^{-1}=1+1/\chi \tau_a$, $t_c^{-1} = 1/\bar D \chi \tau_c$, $f=F\Omega/\chi$, 
$\hat \beta_a = \beta_a\ell/\bar D_a$,   $\hat \chi = \chi\ell^2/\bar D_c\equiv 1/\bar D$, $g= t_c^{-1}\Omega C_{eq}$, 
$d_0=\Omega^2\gamma C_{eq}/\left(k_B\bar T \ell\right)$ (where $\gamma$ is the mean step energy), $\beta_0 = \bar D_c/\left(k_+\ell\right)$, and 
$\bar D = \bar D_c/\bar D_a$. Notice that $\chi \tau_a$ and $\bar D \chi \tau_c$ are the dimensionless reciprocal desorption rates of the precursors and the adatoms, respectively.

The step stiffness and kinetic anisotropies are chosen smooth and periodic \cite{MSL}:
\begin{equation}
\beta_s(\theta) = 1+\epsilon_{s,m} \cos{m\theta},\; -1< \epsilon_{s,m} < 1,
\nonumber
\end{equation}
\begin{equation}
\beta_k(\theta) = 1+\epsilon_{k,m} \cos{(m\theta-m\theta_0)},\;  -1< \epsilon_{k,m} < 1.
\nonumber
\end{equation}
The conditions on the amplitudes $\epsilon_{s,m}, \epsilon_{k,m}$ ensure that the anisotropies are positive and thus no  orientations are ``missing" from the equilibrium and 
kinetic shapes of the step. (The presence of such orientations usually warrants the inclusion of the regularization term in the 
Gibbs-Thomson condition (\ref{GT}) \cite{GDN,K}.)  

Finally, the diffusion tensors have the form \cite{MSL}
\begin{equation}
D_a = \bar D_a \tilde D_a(\psi)\equiv \bar D_a \left( \begin{array}{cc}
d_{11}(\psi) & \epsilon d_{12}(\psi)  \\
\epsilon d_{21}(\psi) & d_{22}(\psi)  \\
\end{array} \right),\quad
D_c = \bar D_c \tilde D_c(\psi),
\nonumber
\end{equation}
where $\tilde D_c(\psi)=\tilde D_a(\psi)$, $\epsilon$ is a small positive parameter (see the next Section), 
$d_{11}=1+\delta\cos{2\psi}, d_{12}=d_{21}=\delta\sin{2\psi}, d_{22}=1-\delta\cos{2\psi}$,
$\delta$ is related to the eigenvalues of the tensor, and $\psi$ is the tensor axes rotation angle. The assumption that the off-diagonal elements
of the tensors are O(1) in $\epsilon$ is consistent with the long-wave expansion presented in the next Section. Clearly, we also assumed that the
diffusion anisotropy is the same for the precursors and adatoms, and the only difference is the magnitudes of the diffusivities $\bar D_a$ and $\bar D_c$. When $\delta=0$, the diffusion is isotropic. When $\delta \neq 0$ and
$\psi$ is the root of $\sin{2\psi}=0$, the diffusion is weakly anisotropic; for other $\psi$ values it is strongly anisotropic. 

The consideration in this paper will be limited to: 
\begin{enumerate}
\item The case $\chi^{-1} < \tau_a$.  This condition means that the time
elapsed prior to the precursor decomposition is less than the time elapsed prior to its desorption;
otherwise, there is no adatoms on a terrace. 
\item The fixed adatom diffusivity $\bar D_c$. The effects of varying the precursor diffusivity $\bar D_a$ will be to some degree investigated. Notice that the variations of $\bar D_a$ and correspondingly, the ratio 
$\bar D$, affect the dimensionless parameters $t_c^{-1},\; \hat \beta_a,\; g$ and $d_0$.
\end{enumerate}

Complementary to the item 2 in the above list, several important remarks regarding dimensionless parameters in our multi-parametric problem are in order. First, through varying the parameter $t_a^{-1}$ in Eq. (\ref{diff_eq_A_ndim}) one can gauge the relative strengths of the precursor's decomposition and desorption. The limit $t_a^{-1} \rightarrow \infty$, or equivalently $\chi \tau_a \rightarrow 0$ corresponds to pure desorption (no decomposition). The opposite limit $t_a^{-1} \rightarrow 1$, or 
$\chi \tau_a \rightarrow \infty$ corresponds to pure decomposition (no desorption). These limits (as well as any finite variations of $t_a^{-1}$) can be achieved either
by varying $\tau_a$ at fixed $\chi$, or vice versa, by varying $\chi$ at fixed $\tau_a$. In the former case $t_a^{-1}$ and $t_a f$ (the far field precursor concentration) are the only parameters that change values, but in the 
latter case also $t_c^{-1}$, $f$, $\hat \chi$ and $g$ change. Similar considerations apply to $t_c^{-1}$. Other reasonable choices of the length and time scales also result in the dependencies of several key dimensionless parameters on $\chi,\; \tau_a,\; \tau_c$ and $\bar D_a$. (See Ref. \cite{ZV_arxiv} for the in-depth (and complicated) discussion of 
the characteristic scales involved in the precursor-mediated growth problem.) 

We do  not make an attempt to fully explore this vast parameter space, rather we choose to demonstrate some key features and trends of the step growth. 
A detailed parametric study of the step dynamics, in particular the impacts of varying strengths of the kinetic and step energy anisotropies, will be published separately.

\setcounter{equation}{0}
\section{Long-wave expansion}
\label{LW}

Our analysis begins with the formal long-wave expansion as in Ref. \cite{GDN}:
\begin{subequations}
\label{Expansions}
\begin{eqnarray}
x & = & \frac{X}{\epsilon},\; t = T_0+\frac{T_2}{\epsilon^2}+\frac{T_4}{\epsilon^4}+...,\; 
A =  A_0(X,z,T_0,T_2,...)+ \epsilon^2 A_2(X,z,T_0,T_2,...)+...,\nonumber \\
C & = & C_0(X,z,T_0,T_2,...)+\epsilon^2 C_2(X,z,T_0,T_2,...)+...,\nonumber
\end{eqnarray}
\end{subequations} 
where $X$ is the long-scale spatial coordinate, $T_0$ is the fast time, $T_2, T_4, ...$ are the slow time variables and $\epsilon \ll 1$ is the small 
and dimensionless expansion parameter \footnote{Although not presented in this paper, the formal linear stability analysis was developed using the simplifying condition of a ``frozen" precursor concentration; this analysis shows conclusively the long-wave instability.}.
Notice that we do not expand the step position $h$, thus $h(X,T_0,T_2,...)$ is $O(0)$ in $\epsilon$, meaning that the
long-wave evolution equation for the step profile that we will derive is strongly nonlinear and thus it is capable of describing large deformations of the step. 
This equation therefore differs from weakly nonlinear equations of Kuramoto-Sivashinsky type derived near the instability threshold, see for example Ref. \cite{BMV,SU}.

Next, we proceed to derive the solutions to the partially coupled diffusion problems (\ref{diff_eq_A_ndim})-(\ref{infty_A_ndim}) and (\ref{diff_eq_C_ndim})-(\ref{infty_C_ndim}) at orders of $\epsilon$ zero, two, and four.
The well-posed evolution equation for the step profile emerges after combining contributions at these orders, similar to Ref. \cite{GDN}.
Without the significant loss of generality, we assume  
$t_c \neq t_a \Leftrightarrow \alpha_1 \neq \alpha_3$ 
(see Appendix for the definitions of $\alpha_1$ and $\alpha_3$).
When this assumption does not hold, the solution is way more complicated,
since a secular terms must be accounted for in the process of solving the ODEs (in the $z$-variable) for $C_0$ and $C_2$.

At the zeroth order we obtain:
\begin{eqnarray}
\hat A_0(z,h) &=& t_af+\alpha_2\mbox{e}^{\alpha_1(h-z)}, \label{A0} \\
\hat C_0(z,h) & = & \alpha_4\mbox{e}^{\alpha_3(h-z)} + q_2\mbox{e}^{\alpha_1(h-z)} + q_1, \label{C0} \\
\bar D^{-1}h_{T_0} & = & -d_{22}\left(\alpha_3\alpha_4+\alpha_1q_2\right).\label{h_T0}
\end{eqnarray}
Expressions for $\alpha_1$ - $\alpha_4$, $q_1$ and $q_2$ in terms of the dimensionless parameters
from Section \ref{Formulation} (and Table \ref{t:ndimpar}) are in Appendix. Notice that the step at this order is straight and
it translates with a constant velocity. 

At the second order the solutions are:


\begin{equation}
\hat A_2 = \left\{\frac{\alpha_2}{4\alpha_1d_{22}}\left(1+2\alpha_1z\right)
\left[d_{11}\left(h_{XX}+\alpha_1h_X^2\right)+
\left(d_{12}+d_{21}\right)\alpha_1h_X\right]+ 
s_2^{(a)}\left(h,h_X,h_X^2,h_{XX}\right)\right\}\mbox{e}^{\alpha_1(h-z)}, 
\nonumber
\end{equation}
\begin{eqnarray}
\hat C_2 = \frac{v\left(h_X,h_X^2,h_{XX}\right)}{2\alpha_3}\left(\frac{1}{2\alpha_3}-z\right)\mbox{e}^{\alpha_3(h-z)}+ 
\frac{u\left(h_X,h_X^2,h_{XX}\right)}{\alpha_1^2-\alpha_3^2}\mbox{e}^{\alpha_1(h-z)}+ \nonumber \\
\frac{w\left(h_X,h_X^2,h_{XX}\right)}{\alpha_1^2-\alpha_3^2}\left(\frac{2\alpha_1}{\alpha_1^2-\alpha_3^2}+z\right)\mbox{e}^{\alpha_1(h-z)}+ 
s_2^{(c)}\left(h,h_X,h_X^2,h_{XX}\right)\mbox{e}^{\alpha_3(h-z)},  
\nonumber
\end{eqnarray}
\begin{eqnarray}
\bar D^{-1}h_{T_2} = \left(\alpha_3\alpha_4+\alpha_1q_2\right)\left\{\left(d_{21}-d_{11}h_X\right)h_X+  
\left(d_{12}+\frac{d_{22}}{2}h_X\right)h_X\right\}+  \nonumber \\
d_{22}\left\{-\alpha_3s_2^{(c)}+\frac{v}{2}\left(h-\frac{1}{2\alpha_3}\right)-\frac{\alpha_1u}{\alpha_1^2-\alpha_3^2}+ 
\frac{w}{\alpha_1^2-\alpha_3^2}\left(1-\alpha_1h-
\frac{2\alpha_1^2}{\alpha_1^2-\alpha_3^2}\right)\right\}. \label{h_T2}
\end{eqnarray}
The functions $s_2^{(a)}$,  $s_2^{(c)}$, $u,\;v,\;w$ are shown in Appendix. Note that $h$ (but not its derivatives) actually cancels from the rhs of Eq. (\ref{h_T2}) after these functions
are substituted.

Solutions in the fourth order are very cumbersome, but they are necessary since the fourth derivative term, $h_{xxxx}$, is needed to
cut-off the short-wavelength instability. (We present only the intermediate compact form of $h_{T_4}$ in Appendix.)  Next, transferring to 
the reference frame moving in the $z$-direction with the velocity $h_{T_0}$, combining derivatives:
\begin{equation}
h_t = \epsilon^{2}h_{T_2} + \epsilon^{4}h_{T_4},
\label{combination}
\end{equation}
and introducing the original variable $x$ (which cancels the powers of $\epsilon$ in Eq. (\ref{combination})) results in the final evolution PDE for the step profile:
\begin{eqnarray}
\bar D^{-1}h_t = \left(p_1^{(2)}+p_1^{(4)}\right)h_{xx}+p_2^{(2)}h_x+p_3^{(2)}h_x^2+ 
p_4^{(4)}h_x^3+p_5^{(4)}h_x^4+p_6^{(4)}h_{xx}h_x+p_7^{(4)}h_{xx}h_x^2+ \nonumber \\  
p_8^{(4)}h_{xx}^2+p_9^{(4)}h_{xxx}+p_{10}^{(4)}h_{xxx}h_x+p_{11}^{(4)}h_{xxxx}, 
\label{h_eq_final}
\end{eqnarray}
where the explicit, final forms of the coefficients are presented in the supplementary materials. The superscript (2) or (4)
refers to the order of the expansion in which the corresponding term emerges. (The $h_{xx}$ term has the contributions
from both the second and fourth orders.) 

The primary facts about Eq. (\ref{h_eq_final}) are as follows:
\begin{enumerate}
\item The linear part of the equation is
\begin{equation}
\bar D^{-1}h_t = \left(p_1^{(2)}+p_1^{(4)}\right)h_{xx}+p_2^{(2)}h_x+p_9^{(4)}h_{xxx}+p_{11}^{(4)}h_{xxxx},
\label{h_eq_final_lin_part}
\end{equation}
where $p_1^{(2)}+p_1^{(4)}, p_{11}^{(4)} < 0$ in the case of a long-wave instability. 
\item When the coefficients of the first and the third derivative terms in Eq. (\ref{h_eq_final_lin_part}) are non-zero, 
the result is the lateral drift (in the $x$-direction) of step the perturbations 
with the speed $\bar D |p_2^{(2)}-p_9^{(4)}|$ (the traveling wave solution). 
\emph{The coefficients $p_2^{(2)}$ and $p_9^{(4)}$ 
vanish when the off-diagonal elements of the diffusion tensors are zero, that is, $d_{12}(\psi)=d_{21}(\psi)=0$. In other words, the diffusion on the lower terrace must be strongly anisotropic for the emergence of the drift.} The drift affects the nonlinear dynamics of the step, as described
in Section \ref{DynAniso}.
In addition to  $p_2^{(2)}$ and $p_9^{(4)}$ vanishing when $d_{12}(\psi)=d_{21}(\psi)=0$, also  the coefficients $p_1^{(4)}$, $p_4^{(4)}$, $p_6^{(4)}$ vanish in this case. Then the nonlinear PDE (\ref{h_eq_final}) simplifies to
\begin{equation}
\bar D^{-1}h_t = p_1^{(2)}h_{xx}+p_3^{(2)}h_x^2+ 
p_5^{(4)}h_x^4+p_7^{(4)}h_{xx}h_x^2+ 
p_8^{(4)}h_{xx}^2+p_{10}^{(4)}h_{xxx}h_x+p_{11}^{(4)}h_{xxxx}.
\label{h_eq_final_simpler}
\end{equation}
Eq. (\ref{h_eq_final_simpler}) most closely resembles the Kuramoto-Sivashinsky (KS)-type equation (4.1) from Ref. \cite{SU}. That equation accounts for weak anisotropy of the line energy, but other anisotropies are not accounted for. 
Eq. (\ref{h_eq_final_simpler}) has the same linear terms as the cited equation, and like that equation it contains the nonlinear terms proportional to $h_x^2$ and $h_{xx}h_x^2$, where the latter term emerges
due to line energy anisotropy. In addition, there is three strongly nonlinear terms $p_5^{(4)}h_x^4,\; p_8^{(4)}h_{xx}^2$ and $p_{10}^{(4)}h_{xxx}h_x$ that are due to the assumed
large deformation of the step ($h(X,T_0,T_2,...)$ is $O(0)$ in $\epsilon$). These terms are not in equation (4.1) of Ref. \cite{SU}, since step deformations are assumed small
in that work.
\end{enumerate}


Eq. (\ref{h_eq_final}) can be written in the conservative form for the slope $q\equiv h_x$:
\begin{equation}
\bar D^{-1}q_t = \left[p_2^{(2)}q+p_3^{(2)}q^2+ 
p_4^{(4)}q^3+p_5^{(4)}q^4+   
p_8^{(4)}q_{x}^2+p_{10}^{(4)}qq_{xx}\right]_x+ \left[\frac{\partial G}{\partial q}+p_9^{(4)}q_x+p_{11}^{(4)}q_{xx}\right]_{xx},
\label{q_eq}
\end{equation}
where $G(q)$ is the double-well ``free energy":
\begin{equation}
G=-m_1q^2+m_2q^3+m_3q^4, \quad
m_1 = - \frac{1}{2}\left(p_1^{(2)}+p_1^{(4)}\right), \; m_2 =  \frac{p_6^{(4)}}{6}, \; m_3 = \frac{p_7^{(4)}}{12}.
\nonumber
\end{equation}
Eq. (\ref{q_eq}) generalizes the convective Cahn-Hilliard equation (CCHE) (51) from Ref. \cite{GDN}. It includes the following additional terms: the slope drift terms $p_2^{(2)}q_x$
and $p_9^{(4)}q_{xxx}$, the convective term $p_4^{(4)}\left(q^3\right)_x$, and the higher-order convective terms $p_8^{(4)}\left(q_{x}^2\right)_x$, $p_{10}^{(4)}\left(qq_{xx}\right)_x$. The coefficients
$p_i^{(j)}$ of these terms vanish when $d_{12}=d_{21}=\beta_0=0$, that is, the terrace diffusion is isotropic and there is no kinetic contribution in the Gibbs-Thomson condition (\ref{GT}).

It follows from Eq. (\ref{h_eq_final}) that the step is linearly unstable with respect to the long-wave perturbations
having wavenumbers $k<k_c=\sqrt{\left(p_1^{(2)}+p_1^{(4)}\right)/p_{11}^{(4)}}$. The maximum perturbation growth rate is attained at $k=k_{max}=k_c/\sqrt{2}$; correspondingly, $\lambda_{max}=2\pi/k_{max}$. 

\setcounter{equation}{0}
\section{Limited study of parametric dependencies}

For the typical values of the dimensional parameters \cite{Gillet}, the translation velocity $h_{T_0}$ of the straight step (Eq. (\ref{h_T0})) is positive, i.e. the step grows, when $q_1>0$. (Notice that $q_1$ is a value of
$\hat C_0$ at $z\rightarrow \infty$, see Eq. (\ref{C0}), and because $\hat C_2, \hat C_4, ...$ are zero 
at $z\rightarrow \infty$, the total dimensional concentration there is $C_{eq}+\Omega q_1$.)
Also, the condition $h_{T_0}>0$ is equivalent to above the threshold precursors concentration at $z\rightarrow \infty$. 
In dimensional units: 
\begin{equation}
z\rightarrow \infty: A_0=F/\left(\tau_a^{-1}+\chi\right) > C_{eq}/(\tau_c\chi). 
\nonumber
\end{equation}
\begin{table}
\begin{center}
\begin{tabular}{|c|c|c|c|c|c|c|c|c|c|c|c|c|c|}
\hline
$t_a^{-1}$ & $f$ & $\hat \beta_a$ & $t_c^{-1}$ & $\hat \chi$ & $g$ & $d_0$ & $\beta_0$ & $\epsilon_{s,m}$ & $\epsilon_{k,m}$ & $\theta_0$ & $\delta$ &  $m$ & $\psi$   \\
\hline
1.02 & 0.02 & 180.5 & 0.5 & 1 & 0.004 & 0.0004 & 0.005 & 0.001 & 0.08 & $0$ & 0, 1/3 & 6 & $\pi/6$\\
\hline
\end{tabular}
\vspace{0.2cm}\\
\caption{The base set of the dimensionless parameters values. These values correspond to $\chi\tau_a =50$, $\chi\tau_c =2$, and  $\bar D =1$.}
\label{t:ndimpar}
\end{center}
\end{table}

\begin{figure}
\vspace{-1.4cm}
\centering
\includegraphics[width=2.0in]{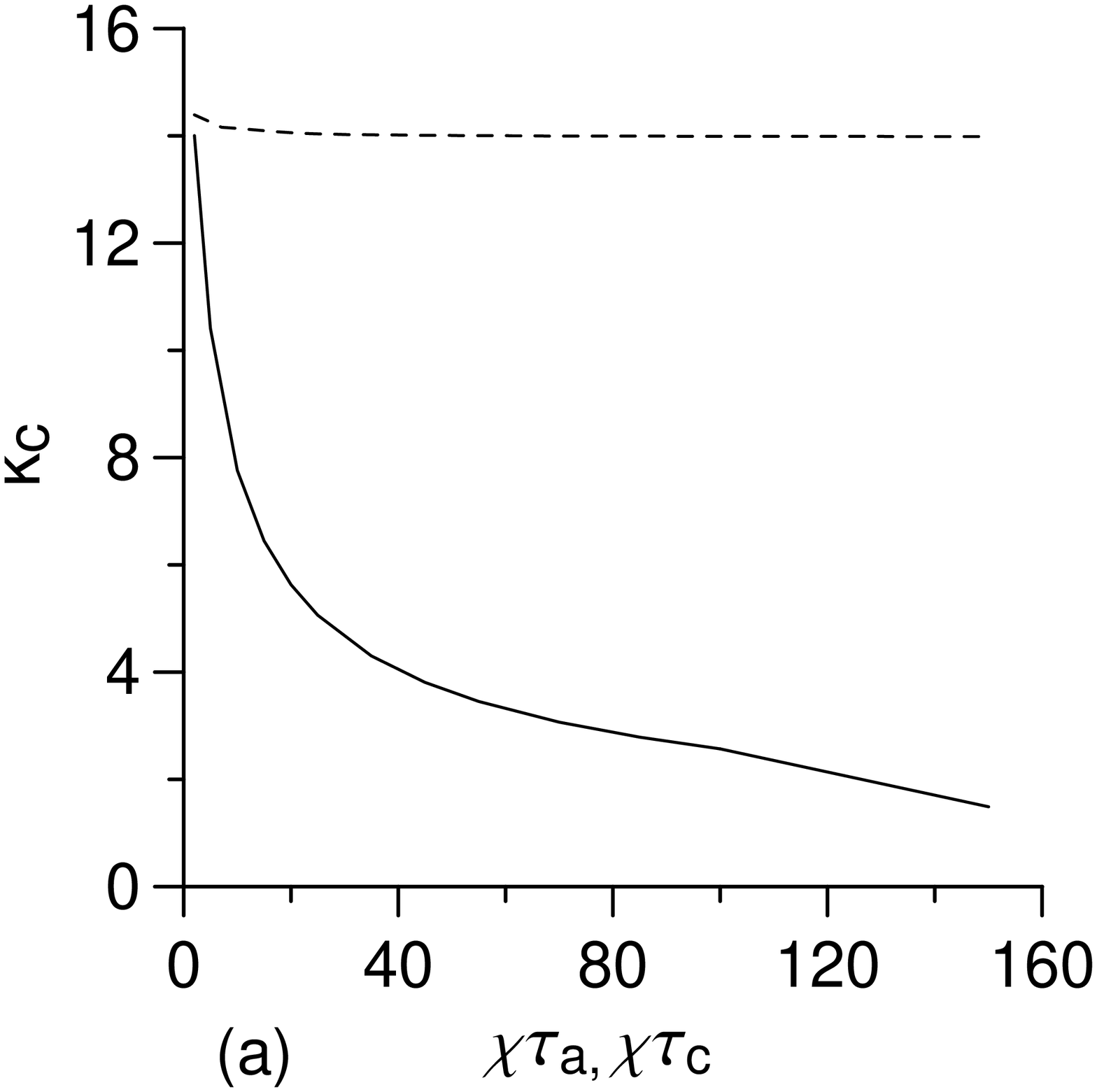}\includegraphics[width=2.0in]{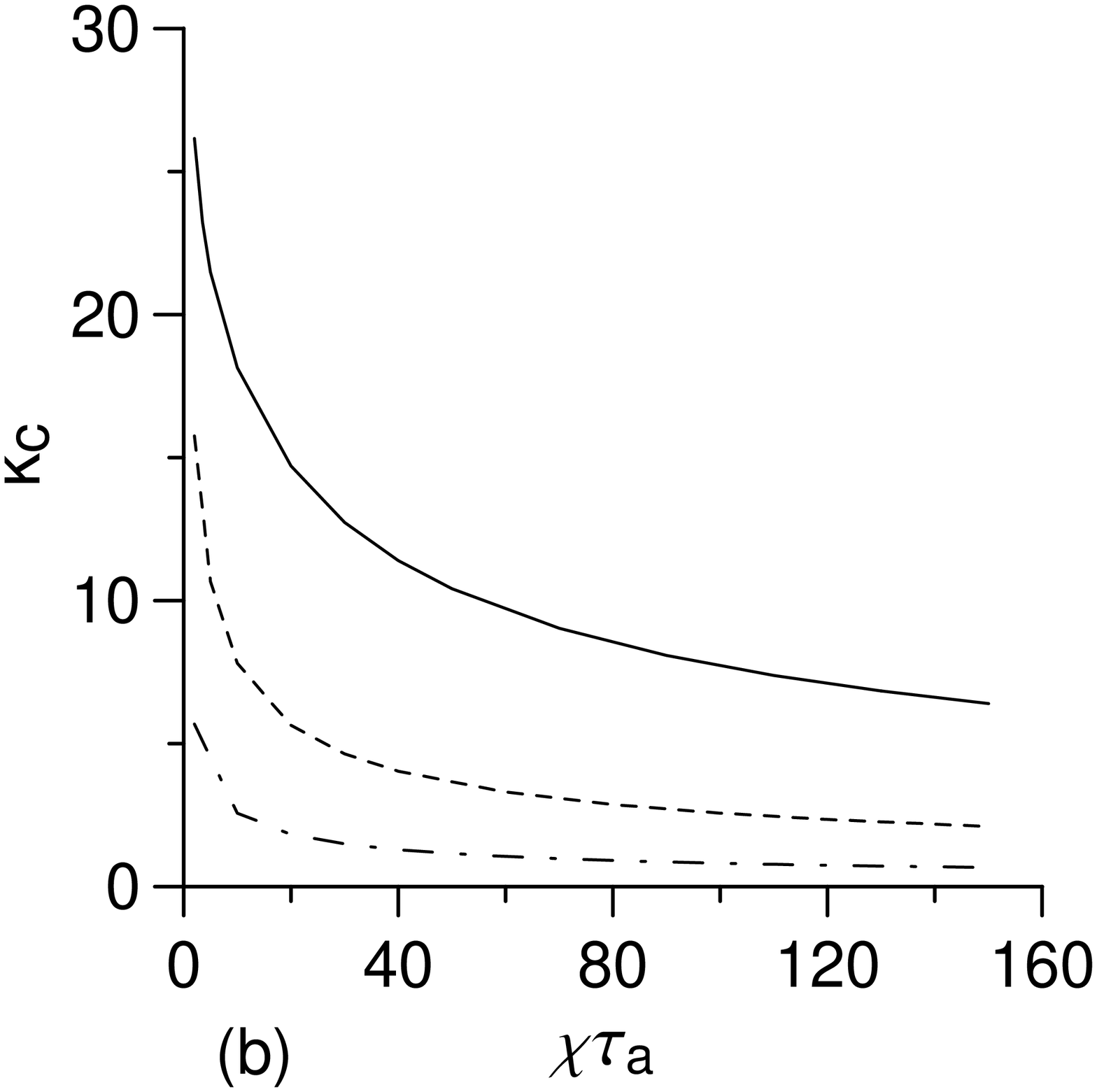} \includegraphics[width=2.0in]{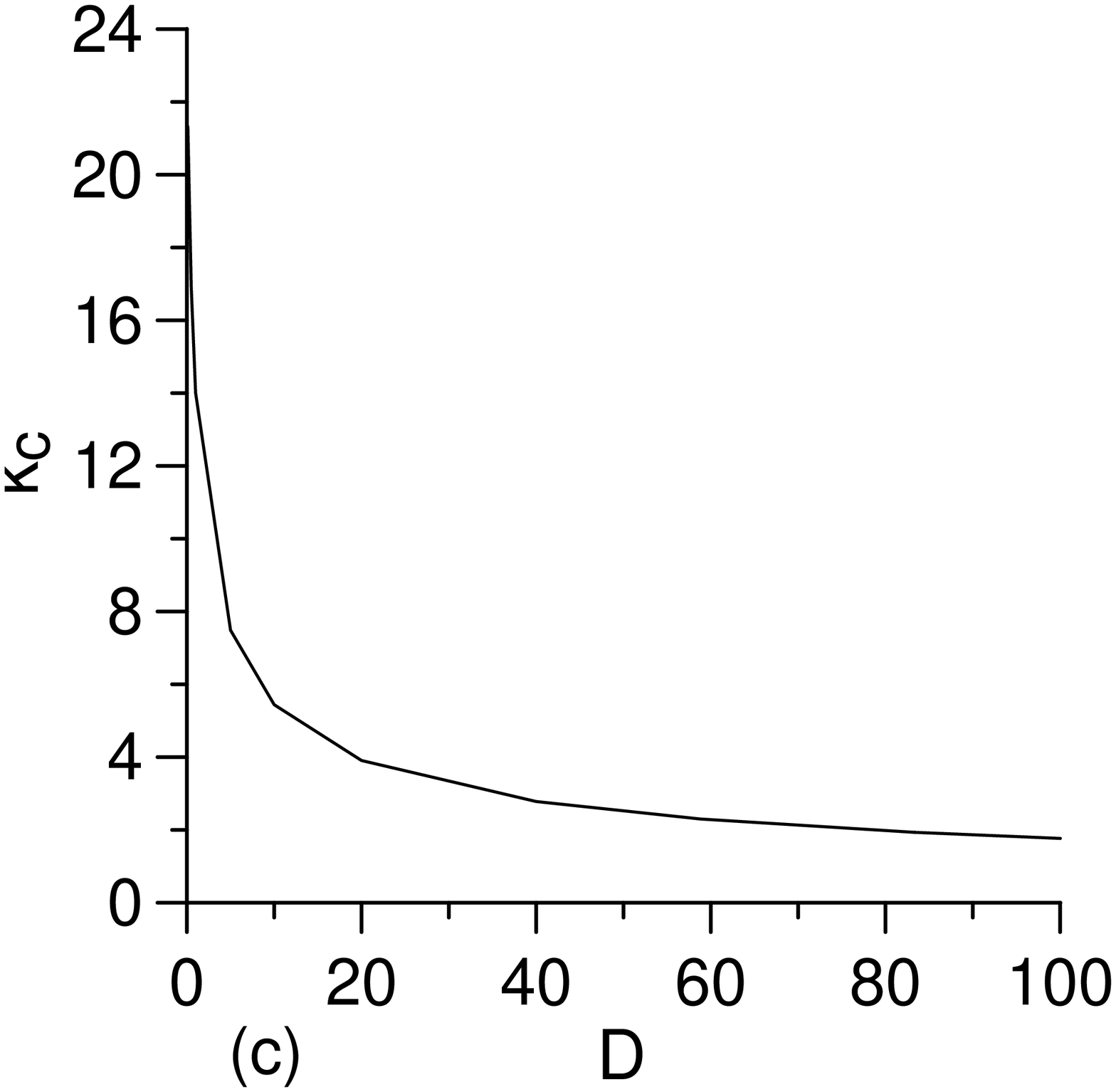}
\vspace{-0.6cm}
\caption{(a) Solid line: $k_c$ vs. $\chi\tau_c$,
where $\chi$ and $\tau_a$ are fixed, such that $\chi\tau_a =50$; $\bar D =1$.  (The latter is equivalent to the constant precursor desorption rate $\left(\chi\tau_a\right)^{-1}=0.02$ and the combined precursors desorption and decomposition rate $t_a^{-1}=1.02$.) Dashed line: $k_c$ vs. $\chi\tau_a$, where $\chi$ and $\tau_c$ are fixed, such that
$\chi\tau_c =2$; $\bar D =1$ $(\equiv t_c^{-1}=1/2)$.
(b) Solid line: $k_c$ vs. $\chi\tau_a$, where $\tau_a$ is fixed and $\tau_c=0.1\tau_a$. Dashed line: Same, but $\tau_c=\tau_a$. Dash-dot line: Same, but $\tau_c=10\tau_a$. 
$\bar D =1$. (c) $k_c$ vs. $\bar D$; $\chi\tau_a =50$, $\chi\tau_c =2$.
}
\label{k_c_vs_ChiTauA_ChiTauC}
\end{figure}

Figures \ref{k_c_vs_ChiTauA_ChiTauC}(a,b,c) show $k_c$ in the isotropic case ($\delta=0$) vs. $\chi\tau_a$,  $\chi\tau_c$, and $\bar D$.  
In Fig. \ref{k_c_vs_ChiTauA_ChiTauC}(a), as the dimensionless adatom 
desorption rate decreases ($\tau_c$ increases at fixed $\bar D$ and  $\chi$) the step becomes more stable, but when the dimensionless precursor desorption rate decreases
($\tau_a$ increases at fixed $\chi$), the stability does not change appreciably. The former echoes the single-species case.
In Fig. \ref{k_c_vs_ChiTauA_ChiTauC}(b), as the 
dimensionless precursor desorption rate decreases ($\chi$ increases at fixed $\tau_a$) the step becomes more stable, the faster so the smaller is the dimensionless adatom 
desorption rate (larger $\bar D \chi \tau_c$). And in Fig. \ref{k_c_vs_ChiTauA_ChiTauC}(c), as $\bar D$ increases (precursor diffusivity $D_a$ decreases) the step becomes more stable.
Combined, the dependencies of $k_c$ on $\chi \tau_a$ shown in Figs. \ref{k_c_vs_ChiTauA_ChiTauC}(a,b) demonstrate that the precursor decomposition impacts step stability significantly more than desorption.
\begin{figure}
\centering
\includegraphics[width=2.3in]{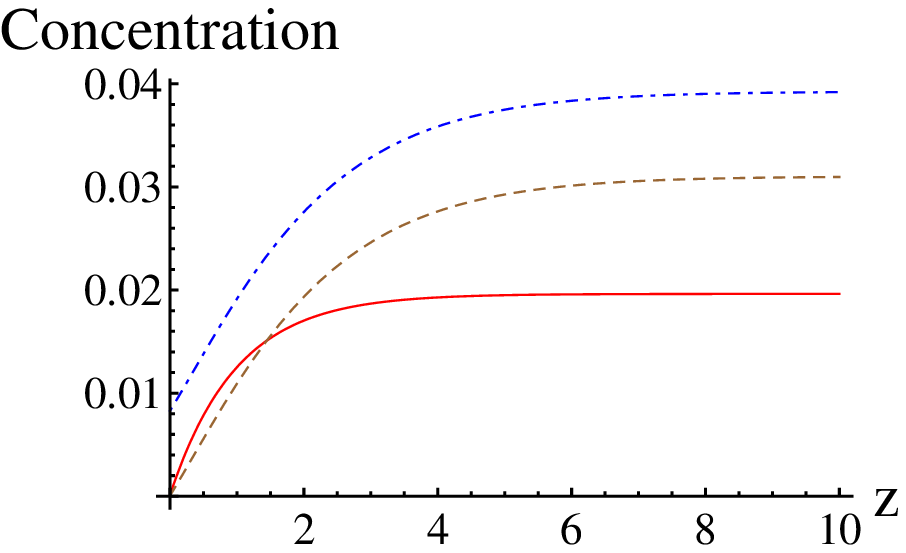}\includegraphics[width=2.3in]{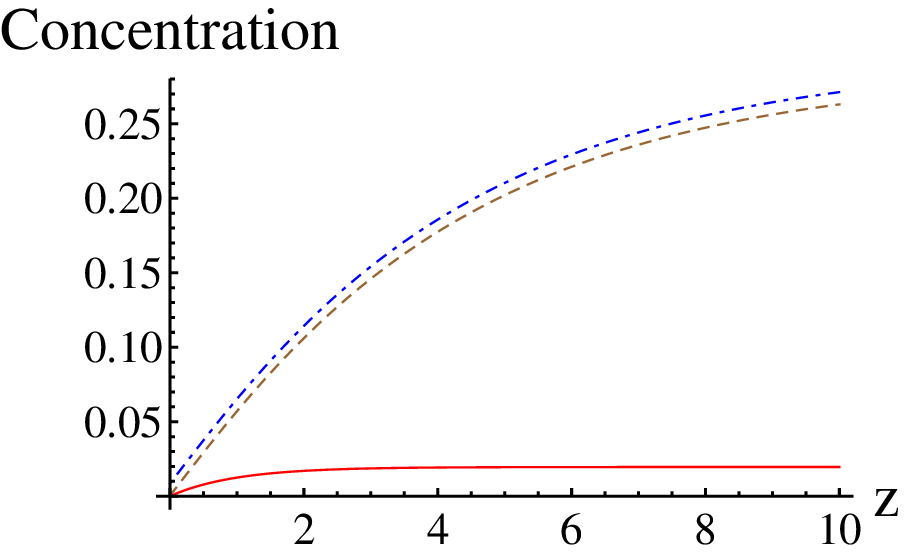}
\caption{(Color online.) The concentration profiles. Solid line: $\hat A_0$, dashed line: $\hat C_0$, dash-dotted line: $\Omega C_{eq}+\hat C_0$. (a) $\chi \tau_a = 50$, $\chi \tau_c =2$,
$\bar D =1$; (b) $\chi \tau_a = 50$, $\chi \tau_c =15$,
$\bar D =1$.}

\label{Con1}
\end{figure}
\begin{figure}
\centering
\includegraphics[width=2.3in]{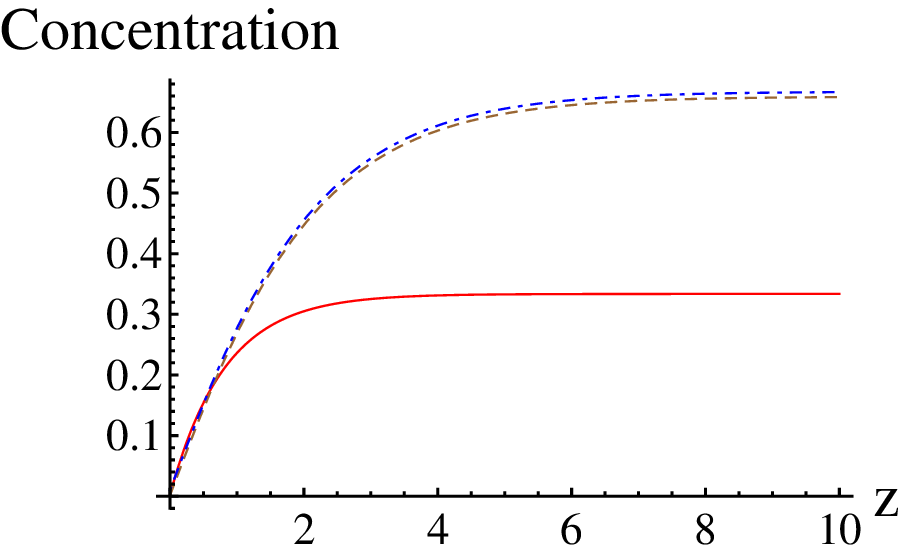}\includegraphics[width=2.3in]{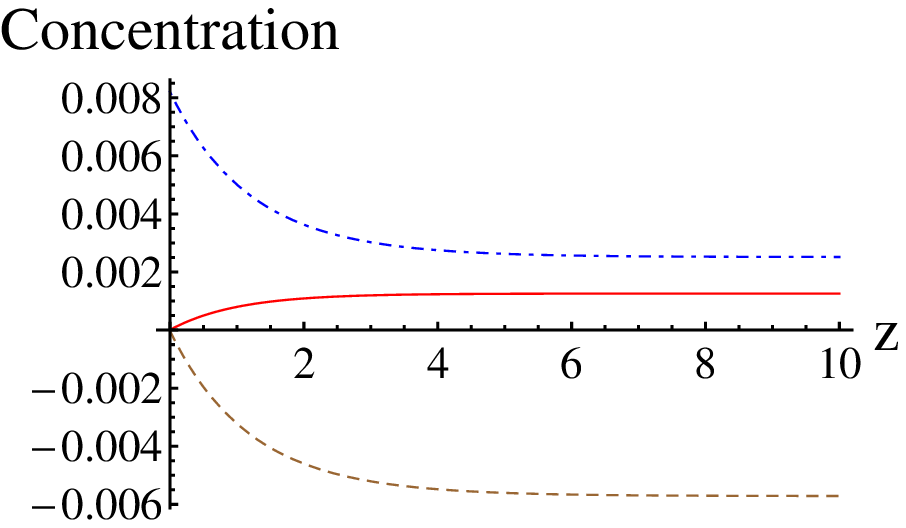}
\caption{(Color online.) The concentration profiles. 
(a) $\chi \tau_a = 2$, $\chi \tau_c =2$,
$\bar D =1$; (b) $\chi \tau_a = 140$, $\chi \tau_c =2$,
$\bar D =1$. The lines have same meaning as in Fig. \ref{Con1}.}
\label{Con2}
\end{figure}
\begin{figure}
\centering
\includegraphics[width=2.3in]{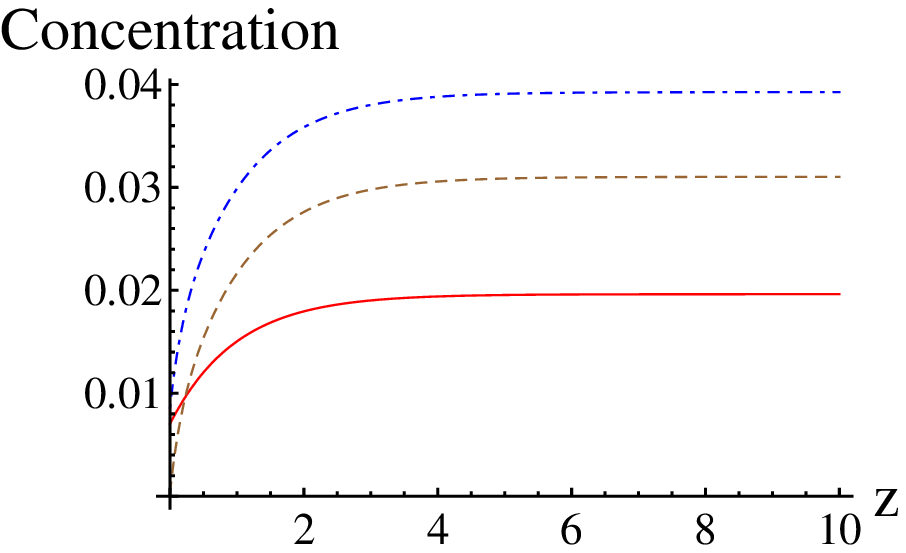}\includegraphics[width=2.3in]{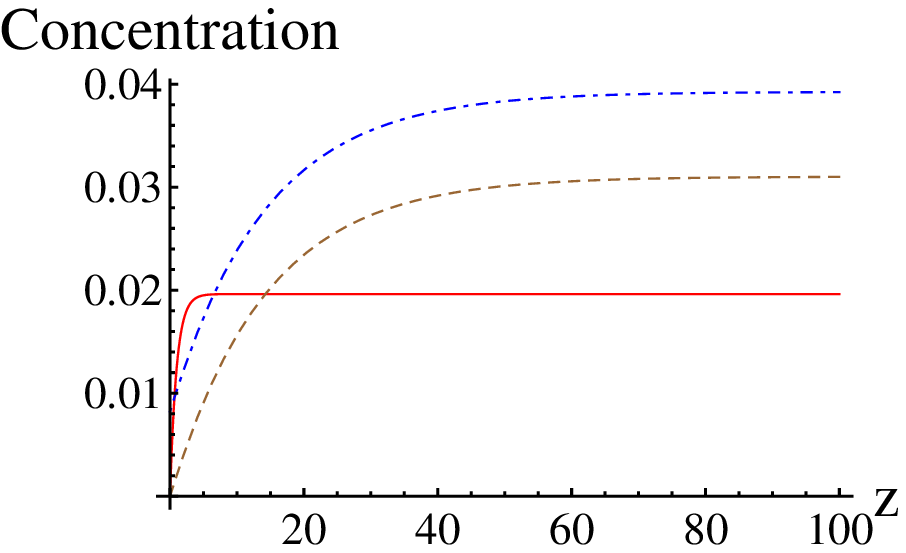}
\caption{(Color online.) The concentration profiles. 
(a) $\chi \tau_a = 50$, $\chi \tau_c =2$,
$\bar D =0.01$; (b) $\chi \tau_a = 50$, $\chi \tau_c =2$,
$\bar D =100$. The lines have same meaning as in Fig. \ref{Con1}.}
\label{Con3}
\end{figure}

Figures \ref{Con1}-\ref{Con3} show the zeroth-order concentration profiles (Eqs. (\ref{A0}) and (\ref{C0}), where we set $h=0$; thus in these figures the step is at $z=0$).   The adatom concentration 
on the terrace
increases roughly linear with the increase of $\chi \tau_c$ (with $\chi \tau_a$ fixed), while the precursors concentration
does not change appreciably (Figures \ref{Con1}(a,b)); this is expected, since $\chi \tau_c$ is the reciprocal dimensionless desorption rate of the adatoms. When instead $\chi \tau_c$ is fixed but $\chi \tau_a$ increases,
both concentrations decrease (Figures \ref{Con2}(a), \ref{Con1}(a), \ref{Con2}(b)); Fig. \ref{Con2}(b) reflects
the situation when the step growth is replaced by evaporation. 
When $\chi \tau_a$ and $\chi \tau_c$ are fixed and $\bar D$ increases, the precursor and adatom concentrations are constant (Figures \ref{Con3}(a), \ref{Con1}(a), \ref{Con3}(b)) -  because the desorption rate $t_a^{-1}$ and the flux $f$ in Eq. (\ref{diff_eq_A_ndim}) do not depend on $\bar D$, and all of the $t_c^{-1},\; \hat \chi$ and $g$ in 
Eq. (\ref{diff_eq_C_ndim}) decrease linearly when $\bar D$ increases. Thus one concludes that varying the precursor diffusivity $\bar D_a$ has no effect on the far-field concentrations on the terrace (see the remark in the end of Section \ref{Formulation} on the connection of $\bar D$ to $\bar D_a$), but it may strongly affect the step dynamics since the parameters at the step, $\hat \beta_a$ and $d_0$, are $\bar D_a$-dependent. In fact, this is confirmed by computations, see Sec. \ref{DynIso}.  

In Sections \ref{DynIso} and \ref{DynAniso} we describe the computations of the step dynamics. The initial condition for the computations is a random, small-amplitude perturbation of the step profile $h(x,0)=1$ on the domain 
$0\le x\le 30\lambda_{max}$, 
with the periodic boundary conditions. 

\setcounter{equation}{0}
\section{Isotropic dynamics ($\delta=0 \Leftrightarrow d_{11}=d_{22}=1, d_{12}=d_{21}=0$)}
\label{DynIso}

In this Section the computations of Eq. (\ref{h_eq_final_simpler}) are described.

\begin{figure}
\vspace{-1.6cm}
\centering
\includegraphics[width=2.5in]{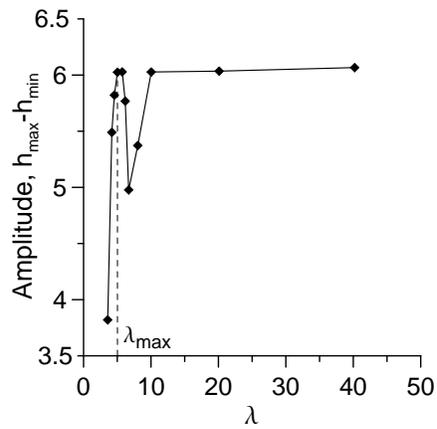}
\vspace{-1.0cm}
\caption{The profile amplitude vs. its wavelength, in the steady-state. Starting from a small-amplitude, single-wavelength cosine curve on a periodic domain, its 
evolution was computed until the steady-state cosine curve-like profile emerged.
$\chi \tau_a = 50$, $\chi \tau_c =2$, $\bar D =100$.}
\label{AmplSat}
\end{figure}

For all parameters values, evolution of the step profile from the random initial condition on the large domain proceeds through coarsening, until it stops and a steady-state profile emerges. 
This scenario is usually termed the interrupted coarsening.  It was argued in Refs. \cite{PM1,PM2} that the signature of \emph{uninterrupted} coarsening is 
the positivity of $dA/d\lambda$ for all $\lambda$, where $A$ is the steady-state profile amplitude and $\lambda$ is profile wavelength,
given that the initial condition is one (unstable) wavelength of the small-amplitude cosine (or sine) curve on the periodic domain.
Indeed, from Fig. \ref{AmplSat} it is clear that this condition does not hold. 

\begin{figure}
\vspace{-0.4cm}
\centering
\includegraphics[width=2.3in]{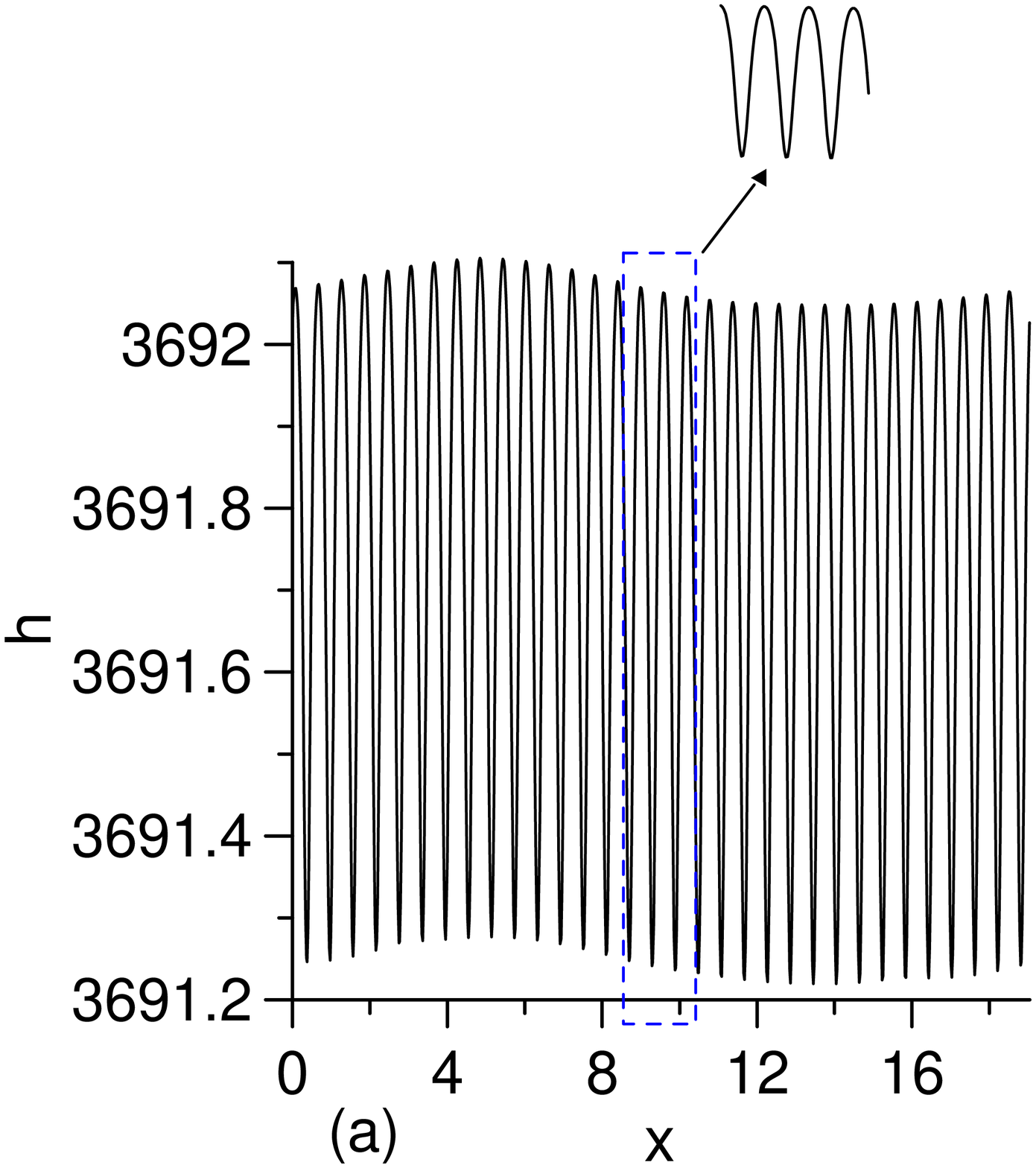}\includegraphics[width=2.3in]{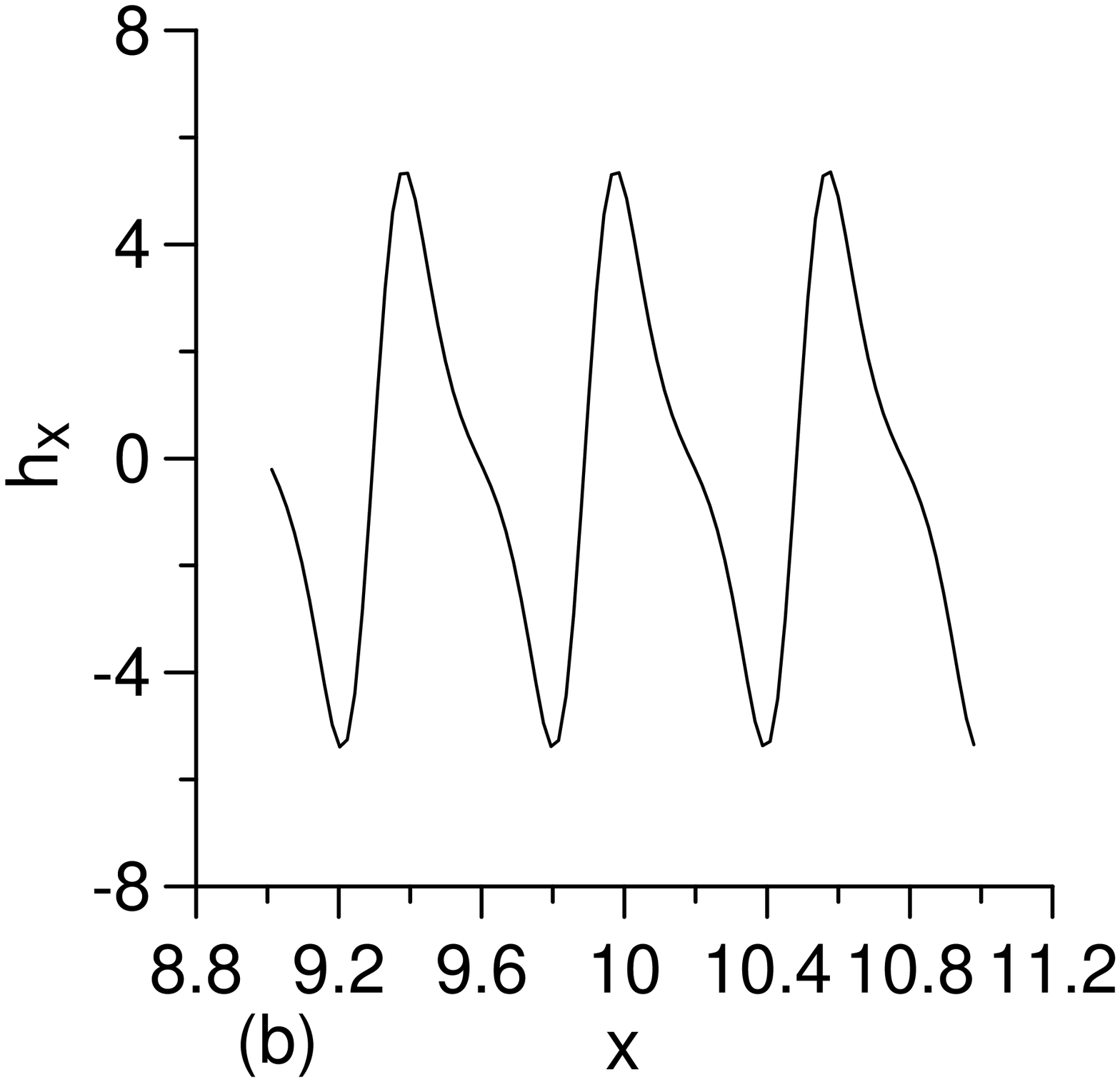}
\vspace{-1.2cm}
\caption{(Color online.) (a) The steady-state step profile. (b) The slope of the steady-state profile (zoom on the interval $9\le x\le 11$, corresponding to inset in panel (a)).
$\chi \tau_a = 50$, $\chi \tau_c =2$, $\bar D =1$.}
\label{ProfileSlopeIso1}
\end{figure}

\begin{figure}
\vspace{-1.4cm}
\centering
\includegraphics[width=2.5in]{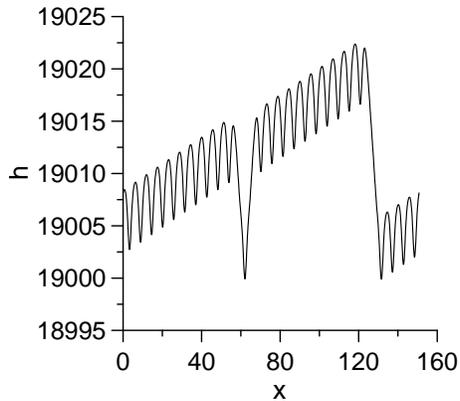}
\vspace{-1.2cm}
\caption{The steady-state step profile. 
$\chi \tau_a = 50$, $\chi \tau_c =2$, $\bar D =100$.}
\label{ProfileSlopeIso2}
\end{figure}

The steady-state step profiles are noticed to be of two types, shown in Figures \ref{ProfileSlopeIso1} and \ref{ProfileSlopeIso2}. The first type is the 
familiar, regular hill-and-valley structure, which may be also described as the periodic faceted structure. (We use the term ``facet" loosely; in fact, the curvature is nowhere zero. Recall that the step energy $\gamma(\theta)$ is a smooth and differentiable function for all step
orientations, and the step stiffness $\beta_s(\theta)>0$ for all $\theta$.) The second type resulted for large $\bar D$ (small $\bar D_a$), and it consists
of the facets bunches. 
For the first type, Fig. \ref{ProfileSlopeIso1}(a) shows the steady-state step profile, and Fig. \ref{ProfileSlopeIso1}(b) shows the corresponding profile slope. 
It can be seen that the profile is asymmetric, with minimas (valleys) being more pointed than the maximas (hills). The second type steady-state step profile
is shown in Fig. \ref{ProfileSlopeIso2}. Here all facets in the periodic computational domain are separated into bunches, with a clear boundary between them. Such unusual bunching
(which to our knowledge has not been previously reported) is attributed to the cumulative effect of the increased precursor decomposition rate at the step, $\hat \beta_a$ (Eq. (\ref{A_cond_h_ndim})), larger adatom concentration at the step through the increased value of $d_0$ (Eq. (\ref{GT_ndim})), and larger step velocity (Eq. (\ref{h_eq_ndim})). These values 
differ by one to two orders of magnitude from the corresponding values in the $\bar D=1$ case shown in Fig. \ref{ProfileSlopeIso1} and Table \ref{t:ndimpar}.

\begin{figure}
\vspace{-1.5cm}
\centering
\includegraphics[width=2.3in]{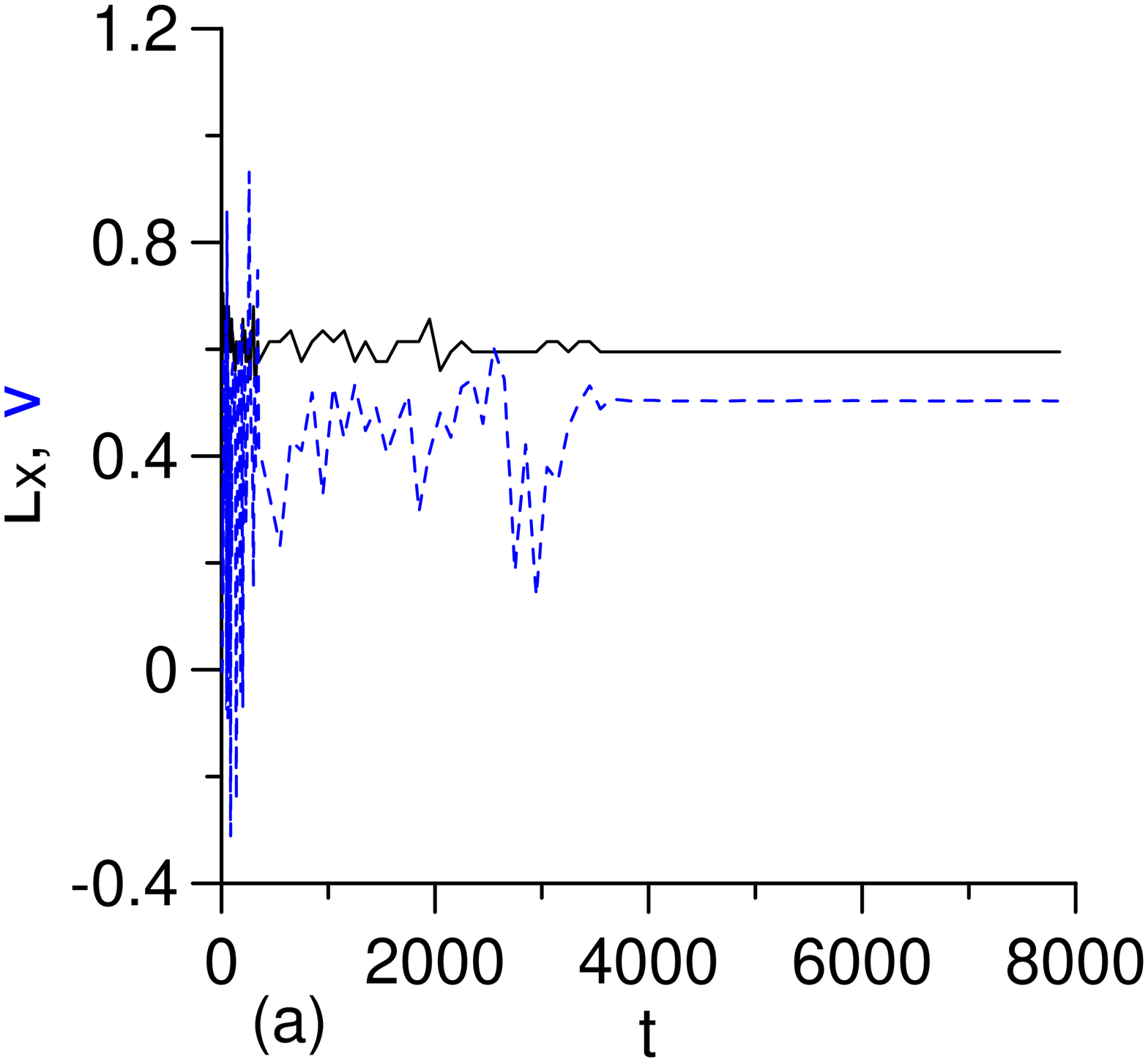}\includegraphics[width=2.3in]{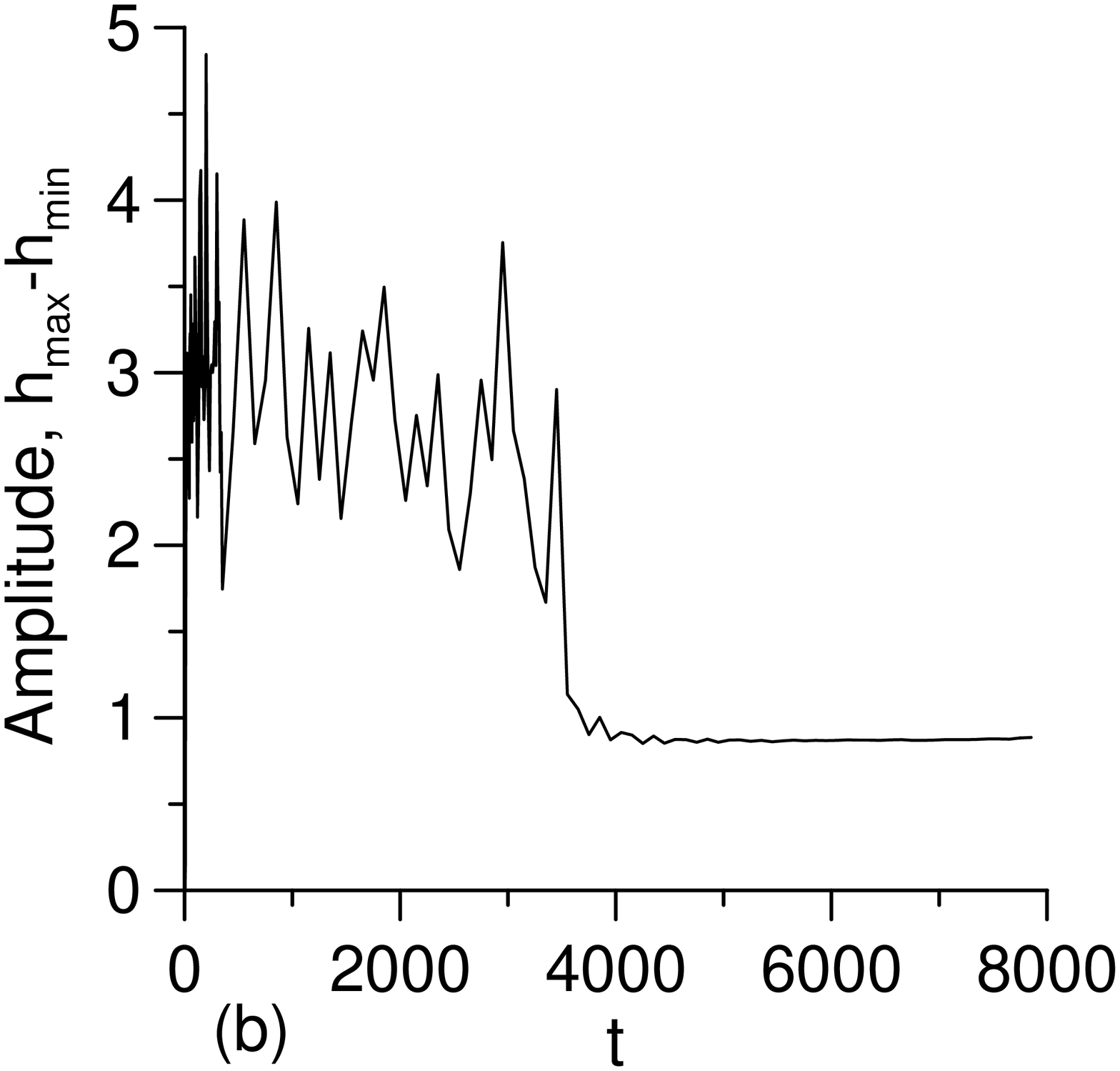}
\vspace{-0.9cm}
\caption{(Color online.) (a) The length scale of the hill-and-valley structure, $L_x$ (solid line) and the step velocity (dashed line) vs. $t$. (b) The  amplitude of the hill-and-valley 
structure vs. $t$. Parameters as in Fig. \ref{ProfileSlopeIso1}.
}
\label{Scale_Velocity_Amplitude}
\end{figure}

Fig. \ref{Scale_Velocity_Amplitude}(a) shows the approach of the characteristic lateral length scale of the profile and the step velocity to the steady-state values. 
The length scale, $L_x$,  is defined as the ratio of the  computational domain length ($30\lambda_{max}$) to the number of valleys. Fig. \ref{Scale_Velocity_Amplitude}(b)
shows the profile amplitude. The steady state emerges at $t=4000 \Leftrightarrow 600$ time units. (This computation was carried up to $t=3.3\times 10^4 \Leftrightarrow 5\times 10^3$ time units, with no change in steady-state values of $L_x$, velocity and amplitude.)

\begin{figure}
\vspace{-1.3cm}
\centering
\includegraphics[width=2.3in]{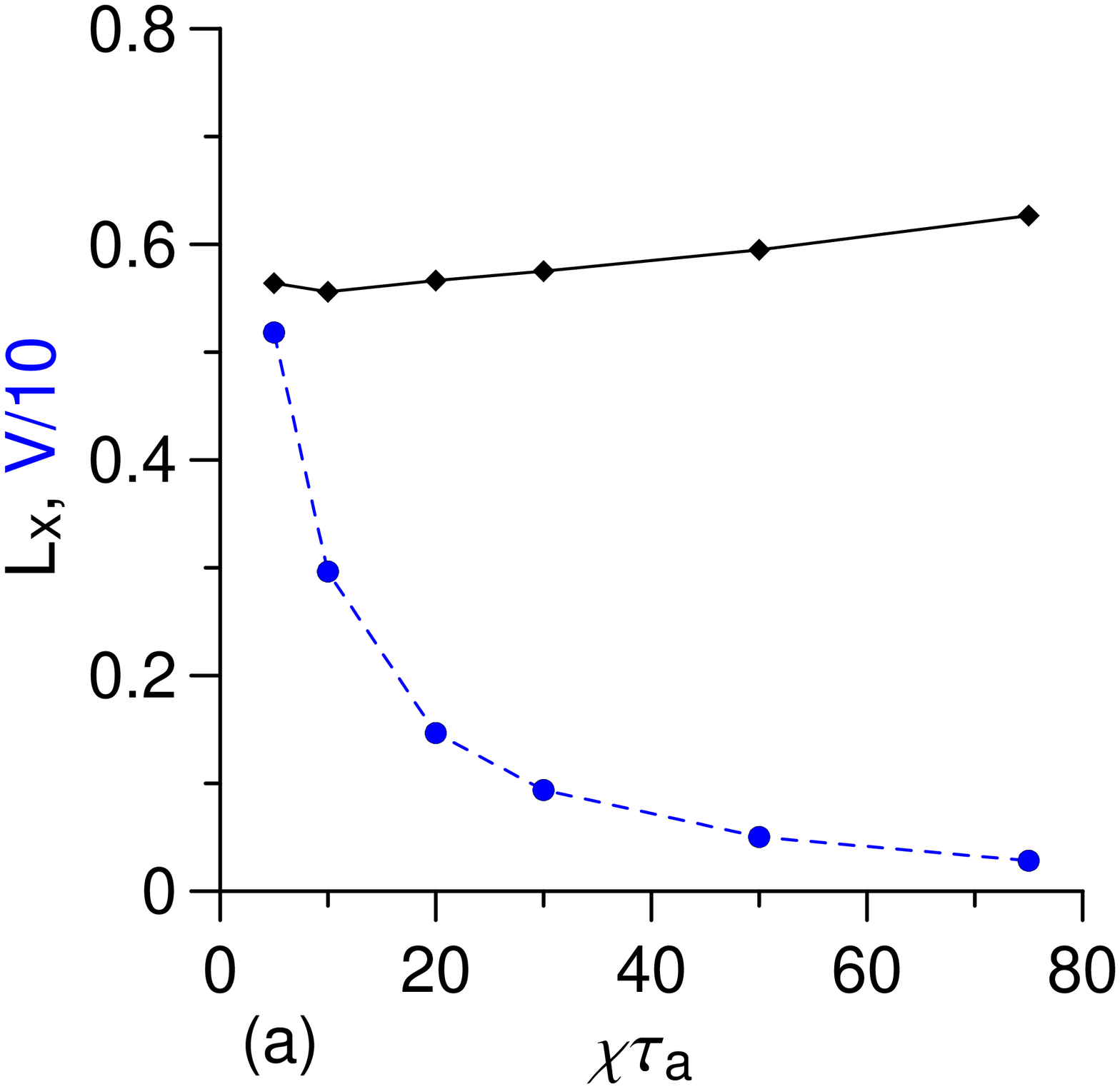}\includegraphics[width=2.3in]{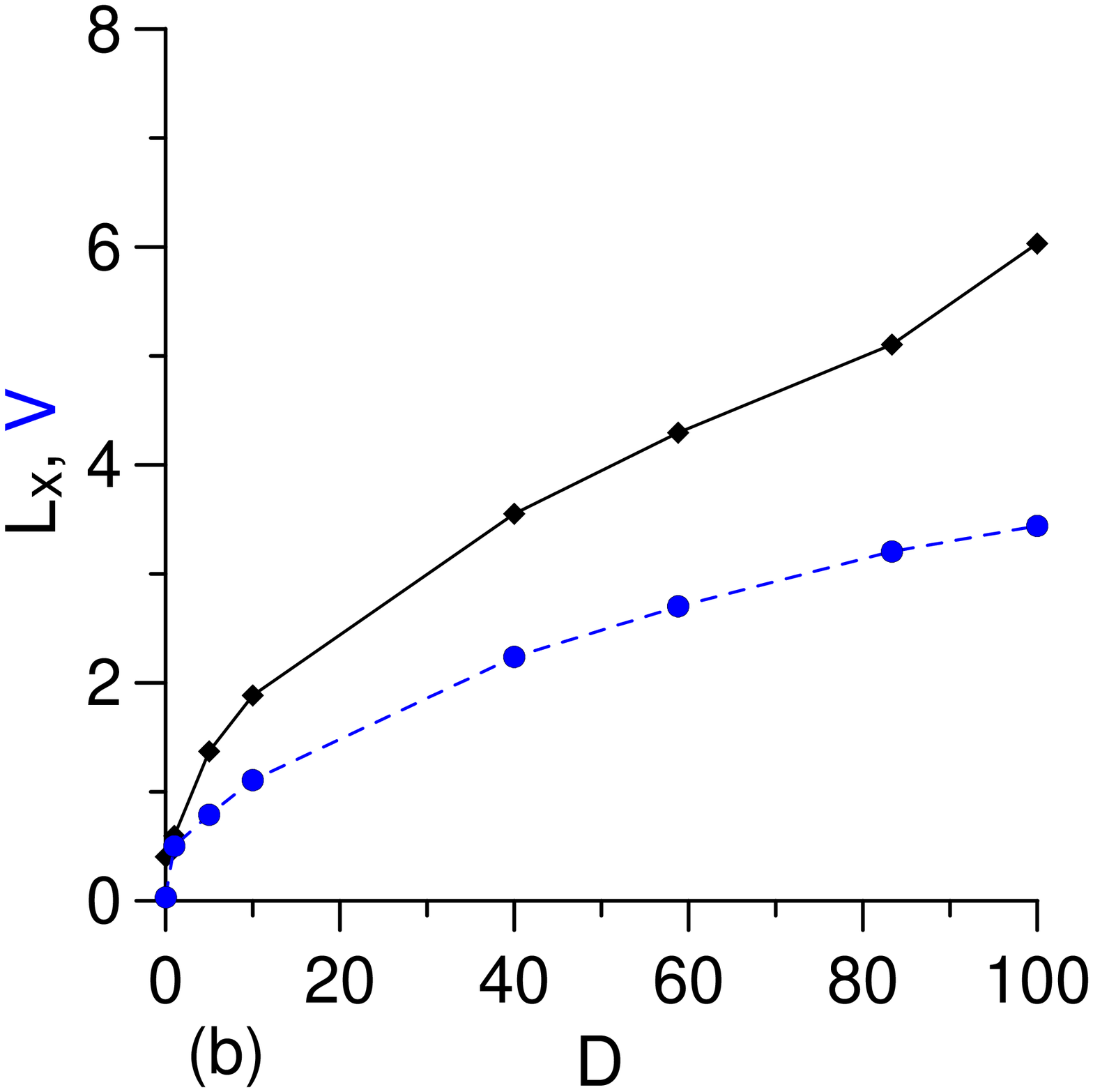}
\vspace{-0.8cm}
\caption{(Color online.) The steady-state length scale of the hill-and-valley structure, $L_x$ (diamonds) and the step velocity (circles). (a) Vs. $\chi \tau_a$, where $\chi$ and 
$\tau_c$ are fixed, such that $\chi \tau_c=2$; $\bar D =1$.
(b) Vs. $\bar D$; 
$\chi \tau_a = 50$, $\chi \tau_c =2$. The curves are only the guides for the eye.}
\label{Scale_Velocity}
\end{figure}

In Figures \ref{Scale_Velocity}(a,b) the steady-state values of the length scale and velocity are plotted vs. $\chi \tau_a$ and $\bar D$. 

%

\setcounter{equation}{0}   
\section{Strongly anisotropic dynamics ($\delta=1/3, \psi=\pi/6 \Leftrightarrow d_{11}=7/6, d_{22}=5/6, d_{12}=d_{21}=1/2\sqrt{3}$)}
\label{DynAniso}

\begin{figure}
\centering
\includegraphics[width=2.0in,angle=-90]{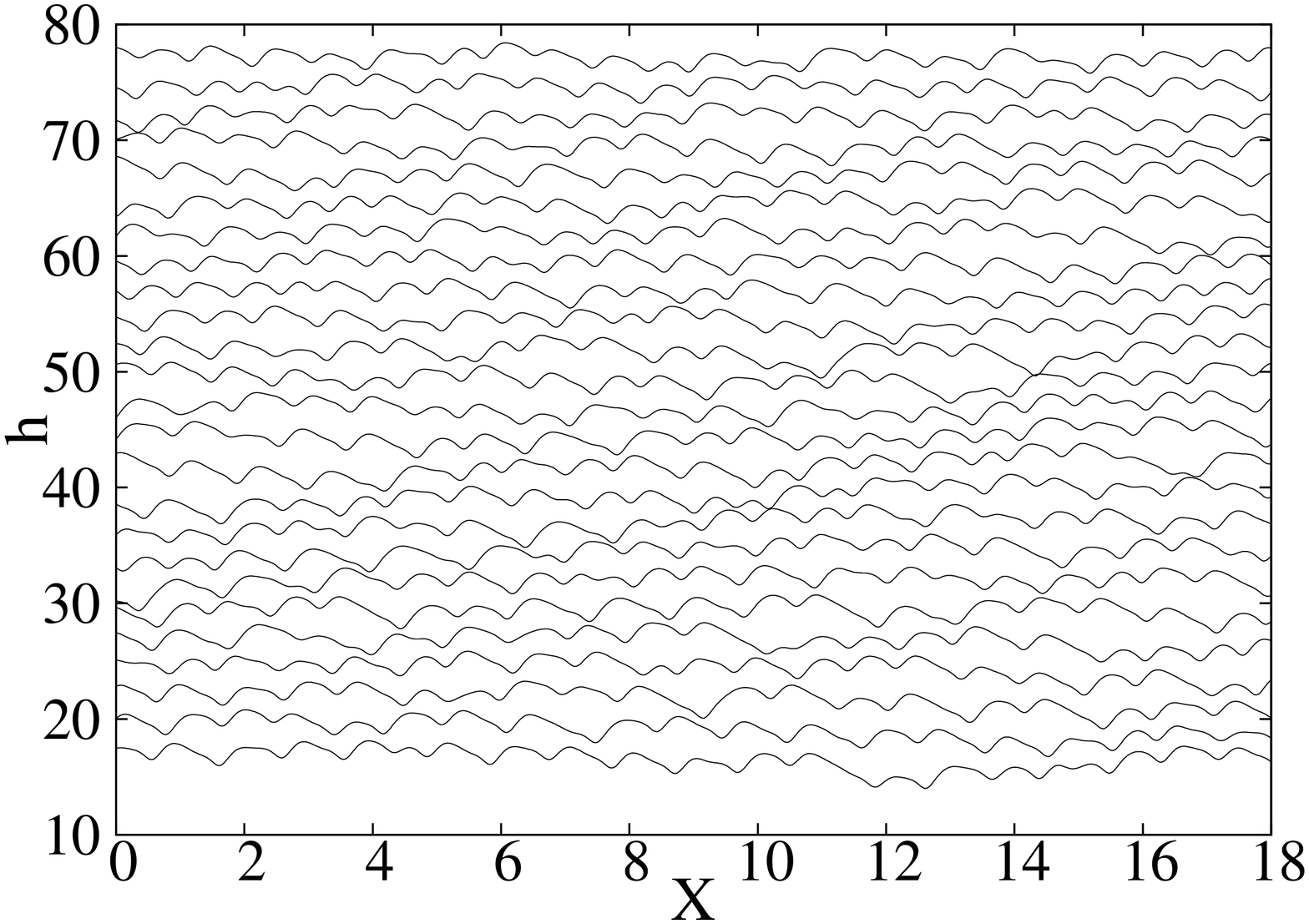}
\caption{Step growth at the strong anisotropy of the terrace diffusion.  
$\chi \tau_a = 50$, $\chi \tau_c =2$, $\bar D =1$. The bottom profile corresponds to $t=80$, the top one to $t=320$. }
\label{ProfileSlopeAniso1}
\end{figure}

\begin{figure}
\vspace{-1.6cm}
\centering
\includegraphics[width=2.3in]{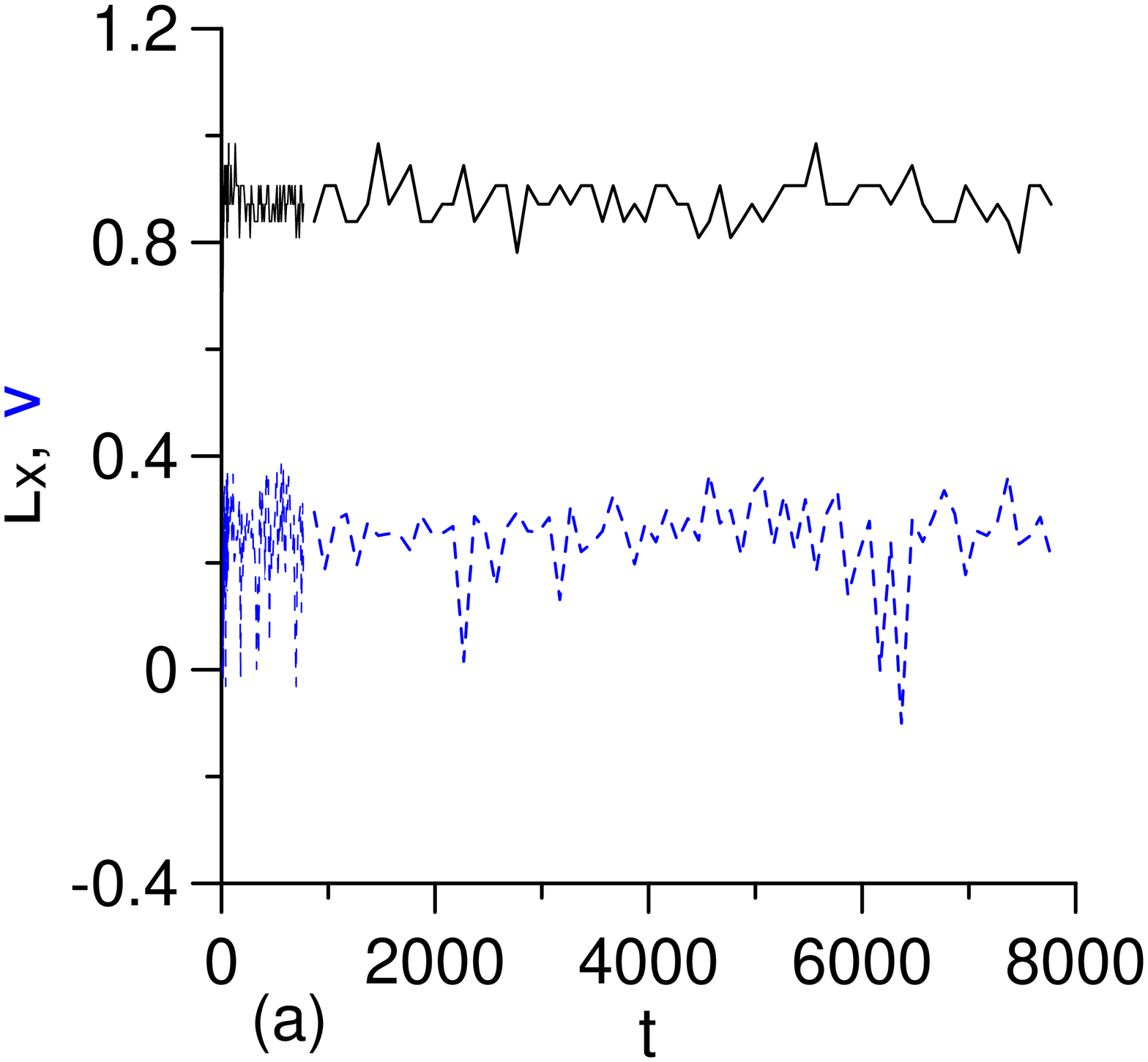}\includegraphics[width=2.3in]{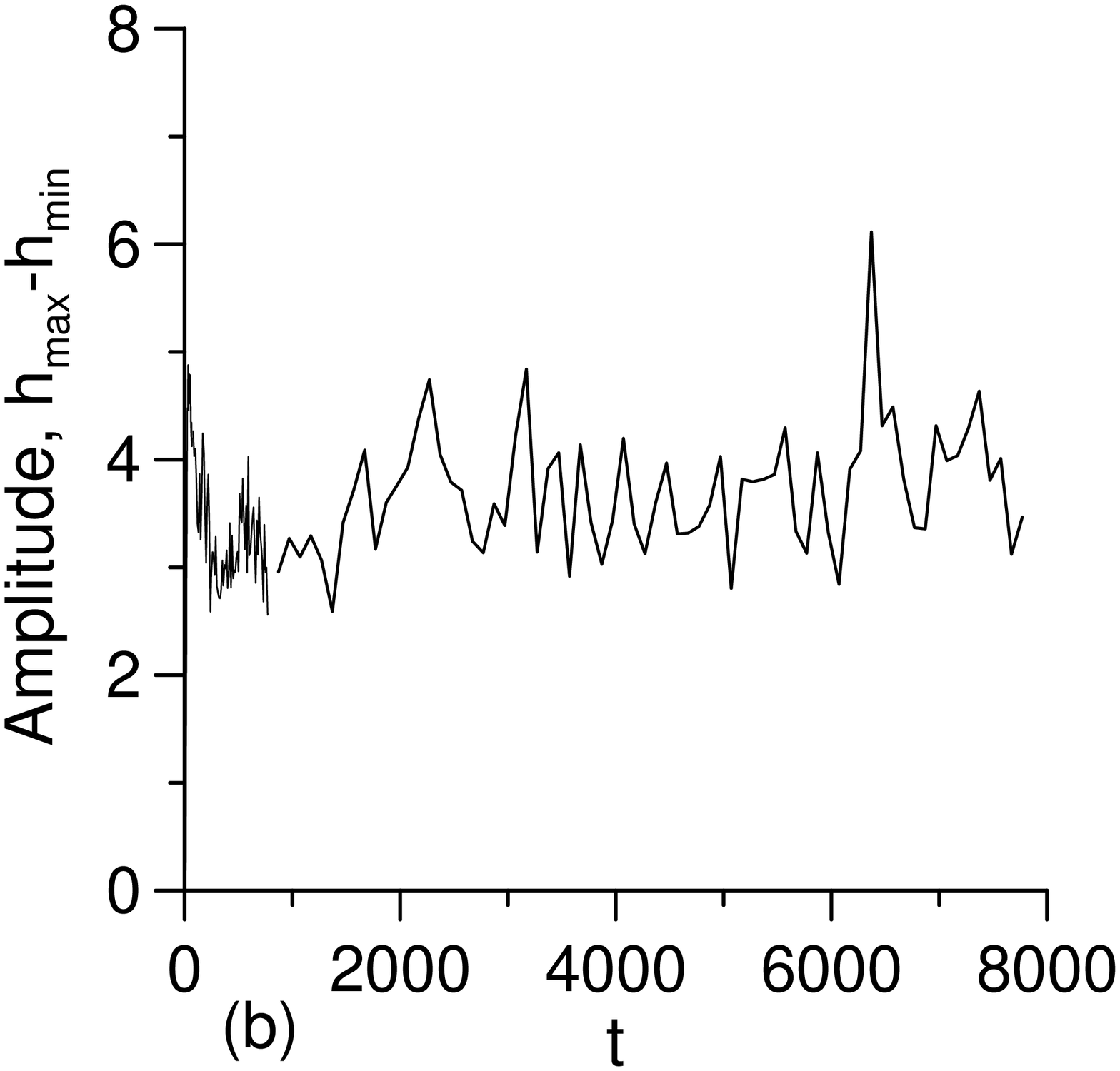}
\vspace{-0.9cm}
\caption{(Color online.) (a) The length scale of step modulations, $L_x$ (solid line) and the step velocity (dashed line) from the computation of step growth shown in Fig. \ref{ProfileSlopeAniso1}. (b) The step profile amplitude.}
\label{Scale_Velocity_Amplitude_Aniso}
\end{figure}

Fig. \ref{ProfileSlopeAniso1} shows some step profiles for $\theta_0=0$ and other parameters as in Fig. \ref{ProfileSlopeIso1}, computed using Eq. (\ref{h_eq_final}). Clearly, there is disorder, and
also there is no highly regular steady-state in the form of a hill-and-valley structure as in Fig. \ref{ProfileSlopeIso1} - the growth as shown in Fig. \ref{ProfileSlopeAniso1} continues 
in the same fashion indefinitely (we computed until $t=2.9\times 10^4$). The disorder is due to a nonlinear traveling wave along the step, triggered and sustained by the $h_x$, $h_{xxx}$, $h_x^3$, and $h_{xx}h_x$ terms in the evolution Eq. (\ref{h_eq_final}). The length scale, velocity and amplitude are shown in Fig. \ref{Scale_Velocity_Amplitude_Aniso}. These quantities oscillate around
the well-defined mean values, with rather large amplitudes (for instance, the step velocity takes on negative values at some times, i.e. the step locally retracts). Also the mean
values of the length scale, velocity, and amplitude themselves are affected by the terrace diffusion anisotropy: both the mean length scale and the amplitude are significantly larger than the 
steady-state, ``isotropic"  values in Figures \ref{Scale_Velocity_Amplitude}(a,b), and the mean velocity is smaller. 


\setcounter{equation}{0}
\section{Conclusions}

The model describing the meandering and growth in the course of a precursor-mediated epitaxy of an isolated, atomically high step on a surface of a thin film  has been analyzed.
A strongly nonlinear evolution PDE for the amplitude of the meander is derived in the long-wave limit and without assuming smallness of the amplitude; this equation may be 
transformed into a convective Cahn-Hilliard-type PDE for the 
meander slope. Computed solutions display an interrupted coarsening and the lateral drift of the meander (a traveling wave),
which affect the important and experimentally measurable parameters, such as the amplitude and velocity. Impacts of the varying precursor diffusivity and the desorption 
rates of the precursors and adatoms on the meander evolution are studied.

In a MOVPE experiments, a step meandering features prominently \cite{GMPV}. The interrupted coarsening and drift have been described previously by Danker \textit{et al.} \cite{DPKM,DPKM1}
and Hauber \textit{et al.} \cite{HV} in the context of models for MBE film growth, where there is a single diffusing species (the adatoms). These effects were attributed 
either to the anisotropy of the line stiffness, or to the terrace diffusion anisotropy. It was noticed, for instance, that the terrace diffusion anisotropy
leads to the tilt of the meander \cite{DPKM} however, in detail such evolution was not studied. In our Eq. (\ref{h_eq_final}) the drift is explicit, its direction is apparent, 
and its speed is the simple expression, unlike in the evolution equations derived in Refs. \cite{DPKM,DPKM1}.

Without the terrace diffusion anisotropy, the step profiles computed from 
our local PDE are similar to those in Ref. \cite{HV}; the latter profiles were computed using the full free-boundary problem. The profiles also resemble
the smoothed versions of the profiles emerging from the analysis of another local PDE, the conserved Kuramoto-Sivashinsky (CKS) equation \cite{FV,GTPB}. The latter PDE is derived
in Ref. \cite{FV} also with the accounting for the terrace diffusion anisotropy.  The connection of these two models deserves, in our view, some exploration in the future.

The computation of the islands growth in the course of MBE and accounting for a full range of the anisotropic effects was done recently in 
Refs. \cite{MSL,MLKMS} using a phase-field model. It is worth noting that we incorporated all such anisotropic effects into a single, closed-form evolution equation that can be 
simplified to fit the MBE setup (by employing the obvious and straightforward recalculations stemming from the omission of the precursors from the model). 

\section*{Acknowledgements}
The author acknowledges support from the grant C-26/628 by the Perm Ministry of Education, Russia.

\appendix
\section{Various dimensionless parameters and functions that appear in the solution expressions presented in section \ref{LW}}
\label{App}

\begin{equation}
\alpha_1 = \sqrt{\frac{t_a^{-1}}{d_{22}}},\;
\alpha_2 = \frac{-\hat \beta_at_af}{\hat \beta_a+\alpha_1d_{22}},\;
\alpha_3 = \sqrt{\frac{t_c^{-1}}{d_{22}}},\;
\alpha_4 = \frac{-q_1-q_2\left(1+\alpha_1\beta_0\left(1+\Upsilon_{k,m}\right)d_{22}\right)}{1+\alpha_3\beta_0\left(1+\Upsilon_{k,m}\right)d_{22}}, \nonumber
\end{equation}
\begin{equation}
\Upsilon_{k,m}=\epsilon_{k,m}\cos{m\theta_0},\; q_1=\frac{\hat \chi t_a f-g}{t_c^{-1}},\; 
q_2 = \frac{\hat \chi\alpha_2}{t_c^{-1}-\alpha_1^2d_{22}}, \nonumber
\end{equation}
\begin{eqnarray}
s_2^{(a)} = \left[\alpha_1\alpha_2h_x\left(d_{21}+d_{12}-d_{11}h_x+\frac{d_{22}}{2}h_x\right)- \right.
\left\{\frac{\alpha_2}{4}\left(2\alpha_1h-1\right)+\frac{\hat \chi \alpha_2}{4\alpha_1d_{22}}\left(2\alpha_1h+1\right)\right\}\times \nonumber \\
\left.\left\{d_{11}\left(h_{xx}+\alpha_1h_x^2\right)+\left(d_{12}+d_{21}\right)\alpha_1h_x\right\}\right]\frac{1}{\hat \chi+\alpha_1d_{22}}, \nonumber 
\end{eqnarray}
\begin{eqnarray}
u = \frac{1}{d_{22}}\left[-\alpha_1d_{11}q_2\left(\alpha_1h_x^2+h_{xx}\right)+\alpha_1^2q_2\left(d_{12}+d_{21}\right)h_x- \right. \nonumber \\
\left.\frac{\alpha_2\hat \chi}{4\alpha_1d_{22}}\left\{d_{11}\left(\alpha_1h_x^2+h_{xx}\right)+\alpha_1\left(d_{12}+d_{21}\right)h_x\right\} -\hat \chi s_2^{(a)}\right],\nonumber 
\end{eqnarray}
\begin{equation}
v = \frac{1}{d_{22}}\left[\alpha_3\alpha_4d_{11}\left(\alpha_3h_x^2+h_{xx}\right)+\alpha_3^2\alpha_4\left(d_{12}+d_{21}\right)h_x\right],\;
w = \frac{-\alpha_2\hat \chi}{2d_{22}^2}\left[d_{11}\left(\alpha_1h_x^2+h_{xx}\right)+\alpha_1\left(d_{12}+d_{21}\right)h_x\right],
\nonumber
\end{equation}
\begin{eqnarray}
s_2^{(c)}=\frac{1}{1+\alpha_3\beta_0\left(1+\Upsilon_{k,m}\right)d_{22}}
\left(-d_0 \left(1+\epsilon_{s,m}\right) h_{xx}+ 
\left(1+\Upsilon_{k,m}\right)\beta_0\left(\left(d_{12}+d_{21}\right)h_x+\left(-d_{11}+\frac{d_{22}}{2}\right) h_x^2\right)\times \right. \nonumber \\
\left(\alpha_3\alpha_4+\alpha_1 q_2\right)+\Upsilon_{k,m}\beta_0 d_{22} h_x^2 \left(\alpha_3\alpha_4+\alpha_1 q_2\right)r_{1,m} 
-\frac{u}{\alpha_1^2-\alpha_3^2}+\frac{v}{4 \alpha_3^2}+\frac{h v}{2 \alpha_3}-\frac{2 \alpha_1 w}{\left(\alpha_1^2-\alpha_3^2\right)^2}-\frac{h w}{\alpha_1^2-\alpha_3^2}+ \nonumber \\
\left(1+\Upsilon_{k,m}\right)\beta_0 d_{22} \left(-\frac{\alpha_1 u}{\alpha_1^2-\alpha_3^2}+\frac{1}{2} \left(h-\frac{1}{2 \alpha_3}\right) v+
\left.\frac{\left(1-\frac{2 \alpha_1^2}{\alpha_1^2-\alpha_3^2}-\alpha_1 h\right) w}{\alpha_1^2-\alpha_3^2}\right)\right),\nonumber 
\end{eqnarray}
\vspace{0.2cm}
\begin{equation}
r_{1,m} = 8,\; \mbox{if}\; m=4;\quad 18,\; \mbox{if}\; m=6, \nonumber
\end{equation}
\vspace{0.2cm}
\begin{eqnarray}
h_{T_4} = \frac{h_X^2}{2}h_{T_2}+\bar D\left[\beta_0\left(1+\Upsilon_{k,m}\right)\left\{d_{21}\hat C_{2X} + 
\left(d_{11}h_X-d_{21}\right)\left(\frac{h_X^2}{2}\hat C_{0X}-\hat C_{2X}\right) + \right. \right. \nonumber \\
\frac{h_X^3}{2}\left(\frac{3h_X}{4}d_{22}+d_{12}\right)\hat C_{0z } - h_X\left(\frac{h_X}{2}d_{22}+d_{12}\right)\hat C_{2z }+ \nonumber \\
\left. d_{22}\hat C_{4z} \right\}-\beta_0 \Upsilon_{k,m}r_{1,m}h_X^2\left\{\left(-d_{11}h_X+d_{21}\right)\hat C_{0X}- \right. \nonumber \\
\left.\left. h_X\left(\frac{h_X}{2}d_{22}+d_{12}\right)\hat C_{0z}+d_{22}\hat C_{2z}\right\} + \beta_0 \Upsilon_{k,m}r_{2,m}h_X^4d_{22}\hat C_{0z}\right], 
\label{h_T4}
\end{eqnarray}
\vspace{0.2cm}
\begin{equation}
r_{2,m} = 16,\; \mbox{if}\; m=4;\quad 66,\; \mbox{if}\; m=6. \nonumber
\end{equation}
In Eq. (\ref{h_T4}) the derivatives of concentrations are understood to be evaluated at the step $z=h$.

\end{document}